\documentclass[showpacs,preprintnumbers,amsmath,amssymb]{revtex4}

\usepackage{graphicx}
\usepackage{dcolumn}
\usepackage{bm}
 
\def\mbi#1{\mbox{\bfseries\itshape #1}} 

\begin{document}

\preprint{APS/123-QED}

\title{Effect of Primordial Magnetic Field on Seeds for Large Scale Structure}

\author{Dai Great Yamazaki$^{1,2}$}
 \homepage{http://th.nao.ac.jp/~yamazaki/}
 \email{yamazaki@th.nao.ac.jp}
\author{Kiyotomo Ichiki$^{3}$}%
\author{Ken-ichi Umezu$^{4}$}
\author{Hidekazu Hanayama$^{1,2}$}%
\affiliation{%
$^{1}$Department of Astronomy, Graduate School of Science, University of Tokyo
7-3-1 Hongo, Bunkyo-ku, Tokyo 113-0033, Japan
}%
\affiliation{%
$^{2}$National Astronomical Observatory, Japan
Mitaka, Tokyo 181-8588, Japan
}%
\affiliation{%
$^{3}$Research Center for the Early Universe, University of Tokyo, 7-3-1
Hongo, Bunkyo-ku, Tokyo 113-0033, Japan
}
\affiliation{%
The Graduate University for Advanced Studies(SOKENDAI), 
Mitaka, Tokyo 181-8588, Japan
$^{4}$
}%

\date{\today}

\begin{abstract}
 Magnetic field plays a very important role in many astronomical
 phenomena at various scales of the universe.  It is no exception in the
 early universe.    
 Since the energy density, pressure, and tension of the primordial
 magnetic field affect gravitational collapses of plasma, the formation
 of seeds for large scale structures should be influenced by them.  Here
 we numerically investigate the effects of stochastic primordial
 magnetic field on the seeds of large scale structures in the universe
 in detail. We found that the amplitude ratio between the density
 spectra with and without PMF ($|P(k)/P_0(k)|$ at $k>0.2$ Mpc$^{-1}$)
 lies between $75\%$ and $130\%$ at present for the range of PMF
 strengths $0.5$ nG $< B_\lambda < 1.0$ nG, depending on the
 spectral index of PMF and the correlation between the matter density
 and the PMF distributions.
\end{abstract}

\maketitle
\section{\label{sec:intro}Introduction}
The possible existence of a primordial magnetic field (PMF) is an 
important consideration in modern cosmology. The origin of PMFs has been
intensively studied \cite{Quashnock89,
Boyanovsky:2002kq,bamba04,Betschart:2003bn,hanayama05,
Takahashi:2005nd,Ashoorioon:2004rs,Ichiki:2006aa}.
Although it seems unlikely to survive an epoch of inflation, it is 
conceivable that large-scale magnetic fields and magnetic inhomogeneities 
could be generated at the end of that era or in subsequent phase 
transitions \cite{R}. Studies of magnetogenesis are partly 
motivated by the need to explain the origin of large-scale magnetic 
fields which are observed in galaxies or in clusters of galaxies \cite{KPZ}. 
If these magnetic fields have the primordial origin, the PMF should have
influenced  a variety of phenomena in the early  
universe \cite{phenome1,Siegel:2006px,semi_num1,Mack02,Lewis04,
yamazaki05a,yamazaki05b,yamazaki06}. 
Several semi-analytic and numerical studies \cite{semi_num1,
Mack02,Lewis04,yamazaki05a,yamazaki05b,yamazaki06} pointed out 
that the effect of the PMF is one of the new physical processes in the 
early universe. 

Large-scale magnetic fields are now considered to be standard
components. Even at high redshift, the existence of dynamically
significant magnetic fields is suggested from observations of 
Faraday rotation associated with high redshift Lyman-$\alpha$ 
absorption systems \cite{KPZ}. 
However their picture in the early universe is not clear and is still 
a matter of debate.
If dynamically significant large-scale magnetic fields were present in 
the early universe, they must have affected the formation and 
evolution of the observed structure, and some signatures of 
their past should be included in their structure and spectrum. 
Indeed, high resolution Faraday rotation maps and the study of 
extragalactic cosmic rays provide direct observational support for this 
point of view \cite{O}.

In the evolution of baryonic and dark matter at large scales, accretion
shocks form in the infalling flows towards the growing nonlinear 
structures such as sheets, filaments and clusters (e.g., \cite{kan94,
ryu98}). 
In fact, some evidence of these accretion shocks has been detected 
in radio relic sources \cite{ens98}. 
The properties of the shocks depend on the power spectrum of the
initial perturbations of baryonic and dark matter on a given scale as
well as the background expansion in a given cosmological model. 
Therefore, if there existed dynamically significant large-scale magnetic
fields in the early universe, the dynamical influence of magnetic fields
on the spectrum of perturbations and thus on flow motions should not
be ignored.
From this point of view, investigations of the physical process in the 
early universe with PMF provide important suggestions for studies of the 
formation and evolution of large scale structures (LSS).

In this paper, we assume the existence of a large-scale PMF and 
analyze the magnetic effects on density inhomogeneities numerically.
Effects of PMF on density fields, especially on cosmic microwave
background (CMB) anisotropies at the
photons' last scattering surface (PLSS), were studied by 
considering the scalar-type component of energy momentum tensor in PMF
\cite{scalar}, or including a new analytical magnetic power spectrum source due to a Lorenz force without previous approximations. 
Here we include both effects consistently and extend their previous
studies by expanding the analysis from the epoch of creation of CMB
anisotropies toward the present epoch. In particular, we consider 
the different correlation between the PMF and the matther power specturm, 
and investigate their
effects on the density fields at large scales in the presence of PMF.

Throughout this paper we fix the cosmological parameters as follows
\cite{Spergel:2003cb}: 
$h=0.71$, $\Omega_b=0.044$, $\Omega_{CDM}=0.226$, $n_s=0.93$, and
$\tau_c=0.10$ in flat Universe models (thus
$\Omega_\lambda=1-\Omega_b-\Omega_{CDM}=0.73$), where $h$ denotes Hubble
parameter in units of 100 km/s/Mpc, $\Omega_b$ and 
$\Omega_m$ are  the baryon and cold dark matter densities in critical
density units, $n_s$ is the spectral index of the primordial scalar
fluctuation, and $\tau_c$ is the optical depth of Compton scattering. 

\section{\label{sec:eff}Cosmological Perturbations with Stochastic
 Primordial Magnetic Field}
\subsection{\label{subsec:cosmoMHD}Cosmological MHD}
Let us consider the PMF created by some effects during the
radiation-dominated epoch.  The energy density of the magnetic field is
treated as a first order perturbation in a flat
Friedmann-Robertson-Walker (FRW) background cosmology.  
The electromagnetic tensor has the usual form
\begin{eqnarray}
{F^\alpha}_\beta=
\left(
    \begin{array}{cccc}
      0    &  E_1  &  E_2  &  E_3 \\
      E_1  &  0    & -B_3  &  B_2 \\
      E_2  &  B_3  &  0    & -B_1 \\
      E_3  & -B_2  &  B_1  &  0  \\
    \end{array}
\right)~,\label{eq_emf_tensor}
\end{eqnarray}
where $E_i$ and $B_i$ are the electric and magnetic fields. 
The energy momentum tensor for electromagnetism is 
\begin{eqnarray}
{T^{\alpha\beta}}_{[\mathrm{EM}]}=\frac{1}{4\pi}\left(F^{\alpha\gamma}F^\beta_\gamma
-\frac{1}{4}g^{\alpha\beta}F_{\gamma\delta}F^{\gamma\delta}\right)\label{eq_ememtensor}.
\end{eqnarray}
The Maxwell stress tensor, consisting of the space-space components of the energy momentum tensor of the electromagnetic field, is  
\begin{eqnarray}
-{T^{ik}}_{[\mathrm{EM}]}=
\sigma^{ik}=  \nonumber \\
\frac{1}{a^2}\frac{1}{4\pi}\left\{E^i E^k+B^i B^k
- \frac{1}{2}\delta^{ik}(E^2+B^2)\right\}\label{eq_MST1}.
\end{eqnarray}
Within the linear approximation, the magnetic field evolves as a stiff
source. Therefore we can discard all back reactions from the magneto
hydrodynamic (MHD) fluid onto the field itself. The conductivity of the
 primordial plasma is very large, and the field is ``frozen-in'' \cite{Mack02}.
 This is a very good approximation for the epochs in which we are interested.
 Furthermore, we can neglect the electric field, $E\sim 0$, and also
 decouple the time evolution of the magnetic field from its spatial
 dependence, i.e., $\mbi{B}(\tau,\mbi{x}) =
 \mbi{B}_0(\mbi{x})/a^2$ for very large scales. In this way we
 obtain the following expressions, 
\begin{eqnarray}
{T^{00}}_{[\mathrm{EM}]}=\frac{B^2}{8\pi a^{6}}~, \label{eq_MST_00} \\
{T^{i0}}_{[\mathrm{EM}]}={T^{0k}}_{[\mathrm{EM}]}=0~, \label{eq_MST_0s} \\
-{T^{ik}}_{[\mathrm{EM}]}=\sigma^{ik}=\frac{1}{8\pi a^{6}}(2B^i B^k -
\delta^{ik}B^2)\label{eq_MST_ss}.
\end{eqnarray}
First and second terms in Eq.(\ref{eq_MST_ss}) are magnetic tension
 and pressure, respectively.
\subsection{Evolution Equations of Cosmological perturbations}
Combining the Einstein equations and linearizing them, we obtain the
perturbation evolution equations. 
In order to consider the effect of the PMF, we should add the
electromagnetic energy momentum tensor to that of standard cosmic fluids, 
\begin{eqnarray}
T^{\alpha\beta}=
{T^{\alpha\beta}}_{[\mathrm{FL}]}+{T^{\alpha\beta}}_{[\mathrm{EM}]}.\label{eq:all_tensor}
\end{eqnarray}
We assume that the baryonic matter is representable as a perfect fluid
and neglect the anisotropic pressure perturbations.  As an initial
condition, we consider adiabatic perturbations and neglect entropy
perturbations initially.
The line element in the conformal synchronous gauge for a flat
Friedmann-Robertson-Walker (FRW) background is given by 
\begin{eqnarray}
ds^2=a^2(\tau)[-d\tau^2 + (\delta_{ij}+h_{ij})dx^idx^j]\label{eq_LECNG},
\end{eqnarray}
where $x^i$ is the spatial coordinate, $a(\tau)$ is scale factor, $\tau$
is the conformal time defined by $d\tau=dt/a(\tau)$, $h_{ij}$ are metric
perturbations around the background spacetime, and the speed of light is set
to unity. 
We will be working in the Fourier space in this paper. We introduce
two fields $h(\mbi{k},\tau)$ and $\eta(\mbi{k},\tau)$ in $k$-space
and write the scalar mode of $h_{ij}$ as a Fourier integral 
\begin{eqnarray}
h_{ij}(\mbi{x},\tau)&=&
	\int d^3
		k e^{i\mbi{k}\cdot\mbi{x}} \nonumber \\
	&\times&
	\left[
		\hat{k}_i\hat{k}_j 
		h(\mbi{k},\tau)
		+6\eta(\mbi{k},\tau)
		\left(
			\hat{k}_i\hat{k}_j-\frac{1}{3}\delta_{ij}
		\right)
	\right]~,
\end{eqnarray}
where $k$ is the wave number in the Fourier space and $\mbi{k}$ is 
$k\hat{\mbi{k}}$ and $\hat{\mbi{k}}$ is a unit vector of wave number.
The linearized perturbation equations are obtained from the Einstein
equations up to first order \cite{scalar,Hu:1997hp,Ma95}: 
\begin{eqnarray}
	k^2\eta-\frac{1}{2}Hh
	&=&
		4\pi Ga^2 \delta T^0_0,\label{eq:scalar_Einstein_00}\\
	k^2\dot{\eta}
	&=&
		4\pi G a^2 ik^j\delta T^0_j,
		\label{eq:scalar_Einstein_0j} \\
	\ddot{h}
	+2H\dot{h}
	-2k^2\eta
	&=&
		8\pi Ga^2\delta T^i_i, \label{eq:scalar_Einstein_ii}\\
	\ddot{h}+6\ddot{\eta}+2H(\dot{h}+6\dot{\eta})-2k^2\eta
	&=& 
		24\pi Ga^2
		\left(
			\hat{k}_i\cdot\hat{k}_j
			-\frac{1}{3}\delta_{ij}
		\right)
		\left(
			\delta T^i_j-\frac{\delta^i_j T^k_k}{3}
		\right)
		\nonumber\\
	&=& 
		24\pi Ga^2
		\left(
			\hat{k}_i\cdot\hat{k}_j
			-\frac{1}{3}\delta_{ij}
		\right)
		\Sigma^i_j
\nonumber\\
	&=& 
		24\pi Ga^2
		\left\{
			Z_\mathrm{[EM:S]}(k)
			-(\rho+p)\sigma
		\right\}
		\label{eq:scalar_Einstein_ij}
\end{eqnarray}
where
\begin{eqnarray}
-\delta T^0_0&=&\delta \rho=\delta \rho_{[\mathrm{FL}]}+\delta
 	\rho_{[\mathrm{EM}]}
 	\label{eq:all_density}~,\\
ik^j \delta T^0_j &=&
	 (\rho+p)v~,\\
\delta T^i_j &=& 
	\delta T^i_{j[\mathrm{FL}]}+\delta T^i_{j[\mathrm{EM}]}~,
\end{eqnarray}
$Z_\mathrm{[EM:S]}(k)$ is a scalar part of the magnetic shear stress, and
$(\rho+p)\sigma$ is a fluid shear stress.
We assume that the PMF $\mbi{B}_0$ is statistically
homogeneous, isotropic and random.  
For such a magnetic field, the power spectrum can be taken as a
power-law $S(k)=<B(k)B^\ast(k)> \propto k^{n} $ \cite{Mack02} where $n$ is the power-law
spectral index of the PMF.
The index $n$ can be either negative or positive depending on the
physical processes of the creation.
From ref.\cite{Mack02}, a two-point correlation function for PMF is defined by
\begin{eqnarray}
\left\langle B^{i}(\mbi{k}) {B^{j}}^*(\mbi{k}')\right\rangle 
	&=&	\frac{(2\pi)^{n+8}}{2k_\lambda^{n+3}}
		\frac{B^2_{\lambda}}{\Gamma\left(\frac{n+3}{2}\right)}
		k^nP^{ij}(k)\delta(\mbi{k}-\mbi{k}'), 
		\ \ k < k_C,
		\label{two_point1} 
\end{eqnarray}
where
\begin{eqnarray}
P^{ij}(k)&=&
	\delta^{ij}-\frac{k{}^{i}k{}^{j}}{k{}^2},\label{project_tensor}
\end{eqnarray}
 $B_\lambda$ is the magnetic comoving mean-field amplitude obtained by smoothing over a Gaussian sphere of comoving radius $\lambda$,
and $k_\lambda = 2\pi/\lambda$ ($\lambda=1$ Mpc in this paper).
The cutoff wave number $k_C$ in the magnetic power
 spectrum is defined by \cite{subramanian98a},
\begin{eqnarray}
k_C^{-5-n}(\tau)=
\left\{
		\begin{array}{rl}
			\frac{B^2_\lambda k_\lambda^{-n-3}}{4\pi(\rho+p)}
			\int^{\tau}_{0}d\tau' 
			\frac{l_{\gamma}}{a},
			& \tau < \tau_\mathrm{dec} \\
			k_C^{-5-n}(\tau_\mathrm{dec}), & \tau > \tau_\mathrm{dec},
		\end{array}
\right.
	\label{eq:CutOff_F}
\end{eqnarray}
where $l_\gamma$ is the mean free path of photons, and $\tau_\mathrm{dec}$ is the time of the decoupling of photons from baryons(see appendix B). 
Evaluating the two-point correlation function of the electromagnetic
stress-energy tensor in the Fourier space, 
we obtain the power spectrum of PMF energy density, Lorenz force and
shear stress as the following (see appendix)
\begin{eqnarray}
|E_{\mathrm{[EM:S]}}(\mbi{k},\tau)|^2\delta(\mbi{k}-\mbi{k}')
=
\frac{1}{(2\pi)^3}
\left\langle
	T(\mbi{k},\tau)_{\mathrm{[EM:S1]}}T^*(\mbi{k}',\tau)_{\mathrm{[EM:S1]}}
\right\rangle
,\label{ED_Souce}
\end{eqnarray}
\begin{eqnarray}
|\Pi_{\mathrm{[EM:S]}}(\mbi{k},\tau)|^2\delta(\mbi{k}-\mbi{k}')
=
\frac{1}{(2\pi)^3}
	\left\langle
	\left(
		T(\mbi{k},\tau)_{\mathrm{[EM:S1]}}
		-T(\mbi{k},\tau)_{\mathrm{[EM:S2]}}
	\right)
	\left(
		T^*(\mbi{k}',\tau)_{\mathrm{[EM:S1]}}
		-T^*(\mbi{k}',\tau)_{\mathrm{[EM:S2]}}
	\right)
	\right\rangle
,\label{LF_Souce}
\end{eqnarray}
and
\begin{eqnarray}
|Z(\mbi{k})_{\mathrm{[EM:S]}}|^2\delta(\mbi{k}-\mbi{k}')
=
\frac{1}{(2\pi)^3}
\left\langle
	\left(
		\frac{2}{3}
		T(\mbi{k},\tau)_{\mathrm{[EM:S1]}}
		-T(\mbi{k},\tau)_{\mathrm{[EM:S2]}}
	\right)
	\left(
		\frac{2}{3}
		T^*(\mbi{k}',\tau)_{\mathrm{[EM:S1]}}
		-T^*(\mbi{k}',\tau)_{\mathrm{[EM:S2]}}
	\right)
\right\rangle
,\label{S_Souce}
\end{eqnarray}
respectively. Explicit expressions for the ensamble averages to evaluate
the above spectra, in the case of power law stochastic magnetic field,
are given as (also see appendix)
\begin{eqnarray}
& &	\langle
		T(\mbi{k},\tau)_{[\mathrm{EM:S1}]}
		T^*(\mbi{k},\tau)_{[\mathrm{EM:S1}]}
	\rangle \nonumber \\
&&=
	\frac{1}{4(2\pi)^7 a^{8}}
	\left\{
		\frac{(2\pi)^{n+8}}{2k_\lambda^{n+3}}
		\frac{B^2_{\lambda}}{\Gamma\left(\frac{n+3}{2}\right)}
	\right\}^2
	\int dk'k'{}^{n+2}
	\left[
	\frac{n^2+4n+1}{kk'n(n+2)(n+4)}
	\left\{
		(k+k')^{n+2}
		-|k-k'|^{n+2}
	\right\}
	\right.
\nonumber\\
&&-
	\frac{1}{k'{}^2n(n+4)}
	\left\{
		|k-k'|^{n+2}
		+|k+k'|^{n+2}
\right\}
+
	\left.
	\frac{k}{k'{}^3n(n+2)(n+4)}
	\left\{
		(k+k')^{n+2}
		-|k-k'|^{n+2}
	\right\}
	\right],
\label{eq:T1T1}
\end{eqnarray}

\begin{eqnarray}
&&	\langle
		T_{[\mathrm{EM:S1}]}(\mbi{k})
		T^*_{[\mathrm{EM:S2}]}(\mbi{k})
	\rangle
	+
	\langle
		T_{[\mathrm{EM:S2}]}(\mbi{k})
		T^*_{[\mathrm{EM:S1}]}(\mbi{k})
	\rangle
\nonumber\\
&&=\frac{1}{(2\pi)^7 a^{8}}
	\left\{
		\frac{(2\pi)^{n+8}}{2k_\lambda^{n+3}}
		\frac{B^2_{\lambda}}{\Gamma\left(\frac{n+3}{2}\right)}
	\right\}^2
    \int dk'k'{}^{n+3}
\left[	
    \frac{1}{(kk')^2n(n+2)}
	\left\{
		(k+k')^{n+3}
	   -|k-k'|^{n+3}
	\right\}
\right.
\nonumber\\ 
&& 	-
	\frac{3}{k^2k'{}^3n(n+2)(n+4)}
	\left\{
		|k-k'|^{n+4}
		+(k+k')^{n+4}
	\right\}
-
	\frac{1}{k^3k'{}^2n(n+2)(n+4)}
	\left\{
		(k+k')^{n+4}
	   -|k-k'|^{n+4}
	\right\}
\nonumber\\ 
&&
\left.
	+\frac{3}{k^3k'{}^4n(n+2)(n+4)(n+6)}
	\left\{
		(k+k')^{n+6}
	   -|k-k'|^{n+6}
	\right\}
\right],
\label{eq:T1T2}
\end{eqnarray}
and
\begin{eqnarray}
&&
\langle 
	T_{[\mathrm{EM:S2}]}(\mbi{k})
	T^*_{[\mathrm{EM:S2}]}(\mbi{k})
\rangle\nonumber\\
&& = 
	\frac{1}{(2\pi)^7 a^{8}}
	\left\{\frac{(2\pi)^{n+8}}{2k_\lambda^{n+3}}
	\frac{B^2_{\lambda}}{\Gamma\left(\frac{n+3}{2}\right)}\right\}^2
	\int dk'
	k'{}^{n+4}
	\frac{4}{(kk')^3n(n+2)(n+4)}
	\left[\frac{}{}
		\left\{
			(k+k')^{n+4}
			-|k-k'|^{n+4}
		\right\}
	\right.
\nonumber\\
&&-
		\frac{3}{(kk')(n+6)}
		\left\{
			|k-k'|^{n+6}
			+(k+k')^{n+6}
		\right\}
+
	\left.
		\frac{3}{(kk')^2(n+6)(n+8)}
		\left\{
			(k+k')^{n+8}
			-|k-k'|^{n+8}
		\right\}
	\right].
\label{eq:T2T2} 
\end{eqnarray}
The evolutions of fluid variables can be obtained by imposing 
the conservation of energy-momentum, which is a consequence of the
Einstein equations 
 \begin{eqnarray}
T^{\mu\nu}{}_{;\mu}=
	\partial_\mu T^{\mu\nu}
	+\Gamma^\nu{}_{\alpha\beta}T^{\alpha\beta}
	+\Gamma^\alpha{}_{\alpha\beta}T^{\nu\beta}
	=0\label{eq:conseration_eqations}.
\end{eqnarray}
This leads the following equations in $k$-space:
\begin{eqnarray}
\dot{\delta}&=&-(1+w)\left(v+\frac{\dot{h}}{2}\right)
			-3H\left(\frac{\delta p}{\delta\rho}-w\right)\delta
			-\frac{3}{8\pi \rho}
			\left\{
				\dot{E}_{\mathrm{[EM:S]}}(\mbi{k},\tau)
				+6HE_{\mathrm{[EM:S]}}(\mbi{k},\tau)
			\right\}~,
			\label{eq:scalar_contiunuity1}\\
\dot{v}&=&-H(1-3w)v
			-\frac{\dot{w}}{1+w}v
			+\frac{\delta p}{\delta\rho}
			\frac{k^2\delta}{1+w}
			-k^2\sigma
			+k^2 \frac{\Pi_{\mathrm{[EM:S]}}(\mbi{k},\tau)}
					 {4\pi \rho} ~,\label{eq:scalar_motion}
\end{eqnarray}
where $w \equiv p/\rho$.
In the continuity and Euler equations for the scalar mode, we can
just add the energy density and pressure of the PMF to the 
energy density and pressure of cosmic fluids, respectively. Since baryon
fluid behaves as a nonrelativistic fluid in the epoch of interest, we
may neglect $w$ and $\delta p_b/\delta\rho_b$, except the acoustic term 
$c_sk^2\delta_b$. Also, the shear stress of baryons is far smaller, and
we can neglect it \cite{Ma95}. Since we concentrate on scalar type
perturbations in this paper, we do not consider the magneto-rotational
instability from the shear stress of the PMF and baryon fluid
\cite{chand61}.

 From equations (\ref{eq:scalar_contiunuity1}) and
 (\ref{eq:scalar_motion}), and by considering Compton interaction
 between baryons and photons, we obtain the same form of the evolution
 equations of photons and baryons as in previous
 works \cite{Hu:1997hp,Ma95,Adams:1996cq},
\begin{eqnarray}
\dot{\delta}_\mathrm{CDM}
 	&=&
 		-\frac{1}{2}\dot{h}~,\label{eq:CDM_rho}\\
\dot{\delta}_{\gamma}
 	&=&
 		-\frac{4}{3}v_{\gamma}-\frac{2}{3}\dot{h}~,\label{eq:photon_rho}\\
\dot{v}_{\gamma}
	&=&
		k^2\left(\frac{1}{4}\delta_{\gamma}-\sigma_{\gamma}\right)
		+an_e\sigma_T(v_\mathrm{b}-v_{\gamma})~,\label{eq:photon_v} \\
\dot\delta_\mathrm{b}
	&=&
		-v_\mathrm{b}-\frac{1}{2}\dot{h}
 			\label{eq:baryon_rho}  \\
\dot{v}_\mathrm{b}
	&=&
			-\frac{\dot{a}}{a}v_\mathrm{b}
 			+c^2_sk^2\delta_\mathrm{b}
 			+\frac{4\bar{\rho}_\gamma}{3\bar{\rho}_\mathrm{b}}
 			an_e\sigma_T(v_{\gamma}-v_\mathrm{b})
			+k^2\frac{\Pi_{\mathrm{[EM:S]}}(\mbi{k},\tau)}{4\pi \rho_b}~,
			\label{eq:baryon_v} 
\end{eqnarray}
where $n_e$ is the free electron density, $\sigma_T$ is the Thomson
scattering cross section, and $\sigma_{\gamma}$ of the second term on
the right hand side of equation (\ref{eq:photon_v}) is the
shear stress of the photon with the PMF. 
Since $n\lesssim 0$ is favored by constraints from the gravitational wave 
background \cite{Caprini:2001nb} and the effect of PMF is not influenced by the time evolution of
the cut off scale $k_C$ for this range of $n$,
we approximately set $E\propto a^{-6}$ in the following analysis.

\section{\label{s:transfer_function}Correlation in power spectra}
In this section, we define a power spectrum function of matter density with 
the PMF.
In the linear approximation, solutions of
eqs.(\ref{eq:CDM_rho})-(\ref{eq:baryon_v}) are divided into those with
and without PMF by Green's function method as, 
\begin{eqnarray}
\delta(k)=\delta_\mathrm{[FL]}(k)+\delta_\mathrm{[PMF]}(k).
\end{eqnarray}
Possible origins of PMF have been studied by many aurthors, however,
we have not reached critical conclusions on the origin of PMF.
Thus we cannot know how the PMF correlates with the primordial density fluctuations.
However, almost all of previous works investigated the effects of PMF on
density perturbations with the assumption that there is no correlation
between them \cite{Tashiro:2005ua}. 
In order to study the PMF effect in a more general manner, 
we have to consider a correlation between the PMF and the primordial density fluctuations.
Therefore we introduce "$s$" to parameterize the correlation between the PMF and the primordial density fluctuations. 
The power spectra of baryon ($P_\mathrm{b}(k)$) and CDM ($P_\mathrm{CDM}(k)$) density with the PMF in the linear approximation are,
\begin{eqnarray}
P_\mathrm{b}(k)&=&
	\left\langle 
		\delta_\mathrm{[b:FL]}(k)
		\delta_\mathrm{[b:FL]}^*(k)
	\right\rangle
	+
	\left\langle 
		\delta_\mathrm{[b:PMF]}(k)
		\delta_\mathrm{[b:PMF]}^*(k)
	\right\rangle 
	 \nonumber\\
	&+&
		2\left\langle 
		\delta_\mathrm{[b:FL]}(k)
		\delta_\mathrm{[b:PMF]}^*(k)
		\right\rangle ,
	 \label{CTFb}\\
P_\mathrm{CDM}(k)&=&
	\left\langle 
		\delta_\mathrm{[CDM:FL]}(k)
		\delta_\mathrm{[CDM:FL]}^*(k)
	\right\rangle
	+
	\left\langle 
		\delta_\mathrm{[CDM:PMF]}(k)
		\delta_\mathrm{[CDM:PMF]}^*(k)
	\right\rangle  \nonumber\\ 
	&+&
		2\left\langle 
		\delta_\mathrm{[CDM:FL]}(k)
		\delta_\mathrm{[CDM:PMF]}^*(k)
		\right\rangle ,
 \label{CTFCDM}
\end{eqnarray}
where, we define the cross correlations as,
\begin{eqnarray}
	\left\langle 
	\delta_\mathrm{[b:FL]}(k)
	\delta_\mathrm{[b:PMF]}^*(k)
	\right\rangle &\equiv& s
	\sqrt{
	        \left\langle 
		\delta_\mathrm{[b:FL]}(k)
		\delta_\mathrm{[b:FL]}^*(k)
	        \right\rangle
	        \left\langle 
		\delta_\mathrm{[b:PMF]}(k)
		\delta_\mathrm{[b:PMF]}^*(k)
	        \right\rangle
		},
\label{crossb}\\
	\left\langle 
	\delta_\mathrm{[CDM:FL]}(k)
	\delta_\mathrm{[CDM:PMF]}^*(k)
	\right\rangle &\equiv& s 
	\sqrt{
	        \left\langle 
		\delta_\mathrm{[CDM:FL]}(k)
		\delta_\mathrm{[CDM:FL]}^*(k)
	        \right\rangle
	        \left\langle 
		\delta_\mathrm{[CDM:PMF]}(k)
		\delta_\mathrm{[CDM:PMF]}^*(k)
	        \right\rangle
		},
\label{crossCDM}
\end{eqnarray}
where $\delta_\alpha$, $\alpha \in (\mathrm{[b:FL]}, \mathrm{[CDM:FL]})$
are baryon and CDM density fluctuations without the PMF respectively, and 
$\delta_\beta$, $\beta \in (\mathrm{[b:PMF]}, \mathrm{[CDM:PMF]})$
are baryon and CDM density fluctuations with the PMF, respectively.  
When $0<s\le 1$, $s=0$, and $-1\le s<0$ on eqs.(\ref{crossb}) and (\ref{crossCDM}), they stand for the positive, no, and negative correlations, respectively.

\begin{figure*}[h]
\includegraphics[width=1.0\textwidth]{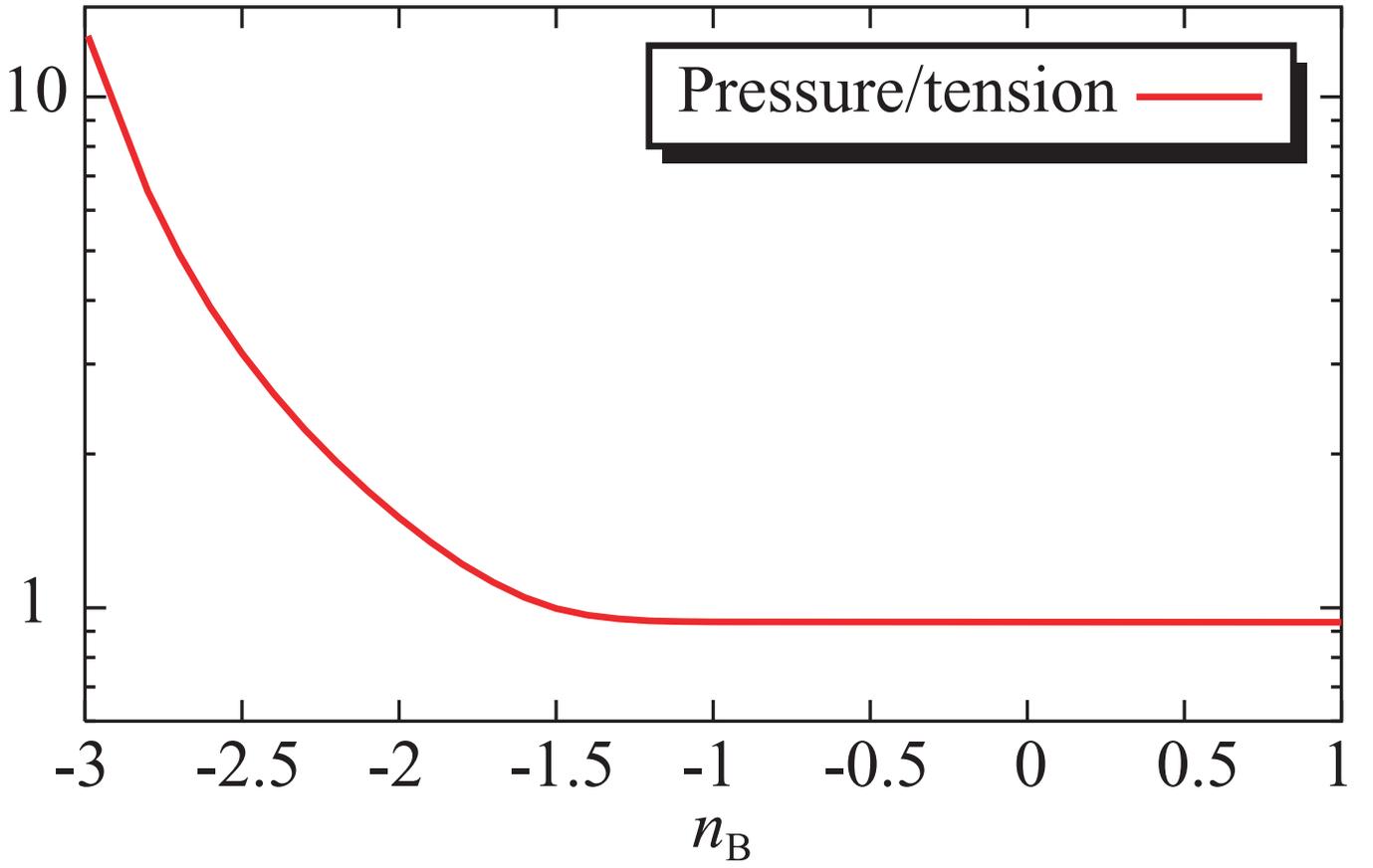}
\caption{\label{fig2}
Ratio of stocahstic PMF pressure and tension sources, 
$
\sqrt{
	\langle
		T_{[\mathrm{EM:S1}]}(\mbi{k})
		T^*_{[\mathrm{EM:S1}]}(\mbi{k})
	\rangle
	/
	\langle
		T_{[\mathrm{EM:S2}]}(\mbi{k})
		T^*_{[\mathrm{EM:S2}]}(\mbi{k})
	\rangle
}
$
, as a function of $n_\mathrm{B}$.
For illustration, a cut off scale $k_C$ is fixed to $k_c = 10$ Mpc$^{-1}$.
The pressure dominates for $n_B < -1.5$,
and the tension dominates for $n_B > -1.5$.
}
\end{figure*}

The square root of power spectrum functions of Lorenz force 
$\Pi_{\mathrm{[EM:S]}}(\mbi{k},\tau)$
in Eq.(\ref{eq:baryon_v}) does not have information about negative or positive,
in other words, there is no information which of the magnetic pressure or tension is dominant, and whether the directions of forces from them are same or different.
However, there must be such information, 
so we should take it into account.
The Lorenz force term in Eq.(\ref{eq:baryon_v}) can be divided into two terms,
 the magnetic pressure and the tension, of which amplitudes are decided by eqs.(\ref{eq:T1T1}) and (\ref{eq:T2T2}), respectively.
Comparing those equations, 
we can decide 
which of them is dominant in the Lorenz force term.
To answer the first question we show in \ref{fig2} which of the magnetic pressure or the tension dominates in the Lorentz force term.
The former dominates when $n < -1.5$,
the latter does when $n > -1.5$. 

As for the second question, we found that the roles of magnetic field 
pressure and tension in the Lorentz force term are different from each 
other for the random primordial magnetic field defined in Eq.(17). 
In other words, the scalar force from magnetic field tension acts on density 
field in the opposite direction in a statistical sense from what magnetic 
field pressure does.  The reason is that, the variance of the
Lorenz force $|\Pi(k)|$ is always smaller than the simple sum of the 
variances of magnetic field pressure and tension, because the cross 
correlation between the two, $T_{\rm [EM:S1]}T^*_{\rm [EM:S2]}$, always 
gives the positive values for all $k$ and $n_s$ considered in this paper.
Since the force field from the magnetic field pressure is directly related 
to the magnetic field energy density distributions, that from the tension
can also be related to the magnetic field energy distributions.
Thus we can decompose the factor as
\begin{eqnarray}
s=s_\mathrm{[LF]}\times s_\mathrm{[DF]},
\end{eqnarray}
where
\begin{eqnarray}
s_\mathrm{[LF]}=
\left\{
		\begin{array}{rl}
			-1, & n < -1.5\ ~\mathrm{(I)},\\
			1, & n > -1.5\ \   ~\mathrm{(II)},
		\end{array}
\right.
	\label{eq:SLF}
\end{eqnarray}
and
\begin{eqnarray}
		\begin{array}{rlccc}
			0 & < & s_\mathrm{[DF]} &\le& 1\ ~\mathrm{(i)},\\
			  &   & s_\mathrm{[DF]} & = & 0\ ~\mathrm{(ii)},\\
			-1&\le& s_\mathrm{[DF]} & < & 0\ ~\mathrm{(iii)}.
		\end{array}\nonumber
\end{eqnarray}
Here $s_\mathrm{[LF]}$ represents
(I) pressure dominant case,
(II) tension dominant case.
On the other hand, $s_\mathrm{[DF]}$ represents
(i) positive correlation between  
the matter and PMF distributions, (ii) no correlation,
and (iii) negative correlation.
Thus if $s<0$, 
it means that 
the the matter and PMF distributions are positively correlated ($s_\mathrm{[DF]}>0$) 
and the PMF pressure dominates in the Lorenz term ($n < -1.5$),
or 
the matter and PMF distributions negatively correlate ($s_\mathrm{[DF]}<0$) 
and
the PMF tension dominates in the Lorenz term ($n > -1.5$).
In these cases the PMF effect acts against the gravitational collapse and 
makes the evolution of density purturbations slower like the gas pressure.
On the other hand, 
if $s>0$, 
the matter and PMF distributions positively correlate ($s_\mathrm{[DF]}>0$) 
and 
the PMF tension dominates in the Lorenz term ($n > -1.5$),
or 
the the matter and PMF distributions negatively correlate
($s_\mathrm{[DF]}<0$)  
and 
the PMF pressure dominates in the Lorenz term ($n < -1.5$). 
In these cases the Lorenz force of PMF accelerates the gravitational collapse.
The above discussion is transparent especially in the
perturbation evolution after decoupling, because $\delta$ does not
ocillate there. 
\section{\label{s:Result}Numerical Results}
In this section, we show our mumerical results.
Since the recent upper limit on PMF amplitude by CMB is $B_\lambda\sim
7.7$ nG at $n_B=-1.5001$\cite{yamazaki06}. 
In order to be consistent with this result, we use the PMF parameter
sets in the following calculations as, $(B_\lambda, n_B)= 
(0.5~\mathrm{nG},-1.0),\ 
(0.5~\mathrm{nG},-2.001),\ 
(1~\mathrm{nG},-1.0)
$,
and
$(1~\mathrm{nG},-2.001)$
\footnote{These values are somewhat ruled out by gravitational wave (GW)
constraint \cite{Caprini:2001nb}. However, because the GW constraint can be applied only for the 
PMF created before the big-bang nucleosynthesis, and because we do not 
specify the origin of the PMF in this paper, we investigate the effect of 
PMF on density perturbations independently of the GW constraint}.

\subsection{Before decoupling}
The effect of PMF on CDM is much smaller than that on baryon before
decoupling because the PMF cannot affect CDM directly. Moreover, 
the density of baryons oscillates with that of photons and their
gravitational effects on CDM are very small.  So we only consider the
PMF effect on baryons here. 

Since the PMF can both effectively increase or decrease the gas pressure, 
the PMF affects the frequency of acoustic oscillations before the baryon
fluid decouples from the photon fluid. 
This is important when considering the CMB anisotropies.
Since the photon tightly couples with the baryon before the PLSS, it is
natural that the frequency of oscillation of the photon density is indirectly
increased or decreased by the presence of PMF.  In this way the PMF affects the CMB
photons which we can observe at present (see \cite{yamazaki05b,yamazaki06}).
\subsection{After decoupling}
\begin{figure*}[h]
\includegraphics{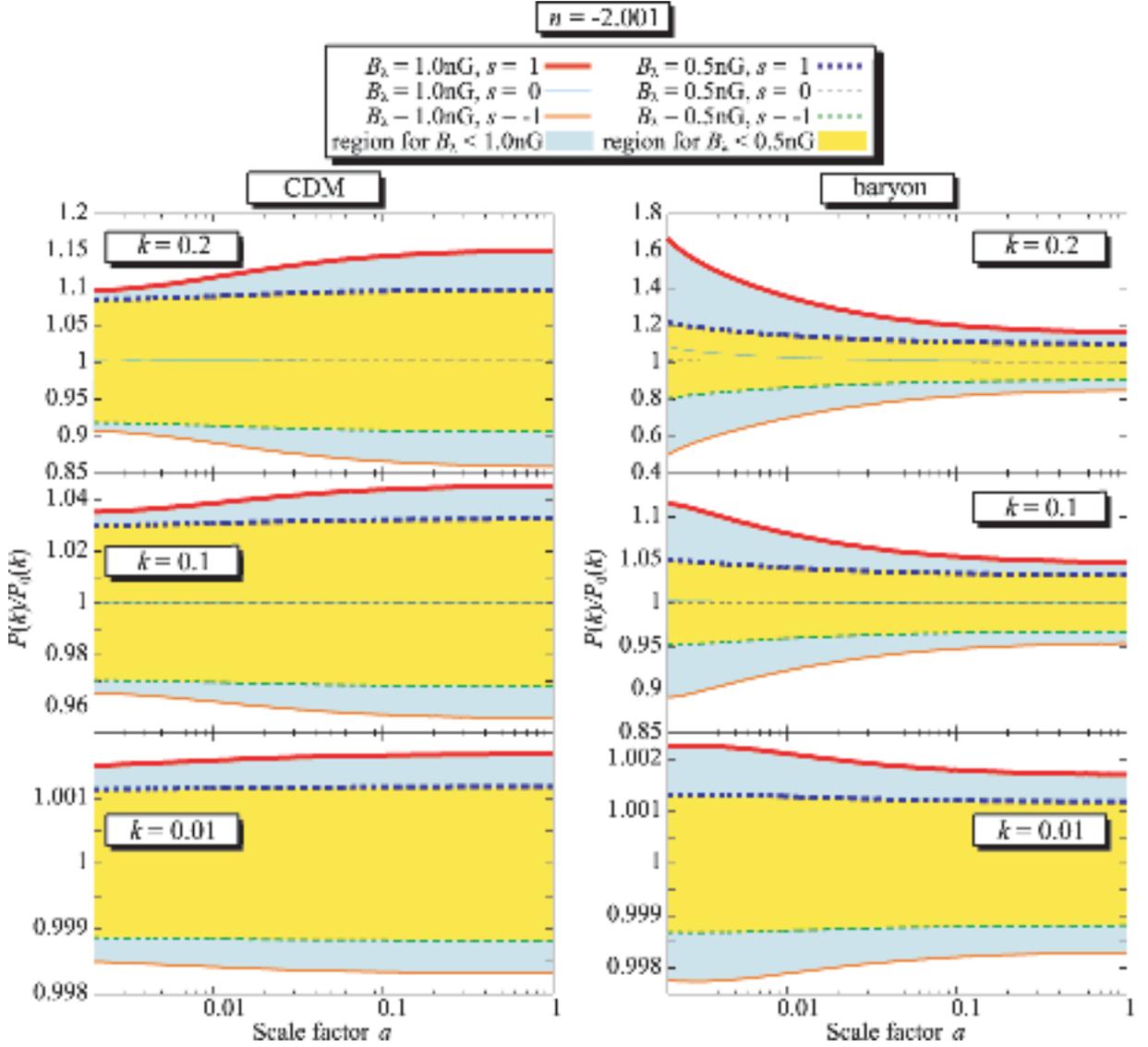}
\caption{\label{fig3}
Differences of transfer functions of CDM and baryon with ($P(k)$) and
 without ($P_0(k)$) the PMF, normalized by $P_0(k)$, with various
 strength of magnetic fields as functions of scale factor $a$ at each wave number $k$ as indicated.
The curves in right and left panels correspond to the differences in CDM
  and baryon, respectively.
The red thick, light-blue thin, orange normal, blue dotted bold, black
 dotted thin, and gree dotted middle curves in each both side panels
 show the differences for parameters
$(B_\lambda ,s)=$ 
(1.0nG,1),
(1.0nG,0),
(1.0nG,-1),
(0.5nG,1),
(0.5nG,0), and
(0.5nG,-1),
respectively.
In all figures, the power spectral index of the PMF is fixed to $n=$ -2.001.
The curves with $s=0$ in all panels are slightly above unity, although
it is difficult to read off from the figure.}
\end{figure*}

\begin{figure*}[h]
\includegraphics{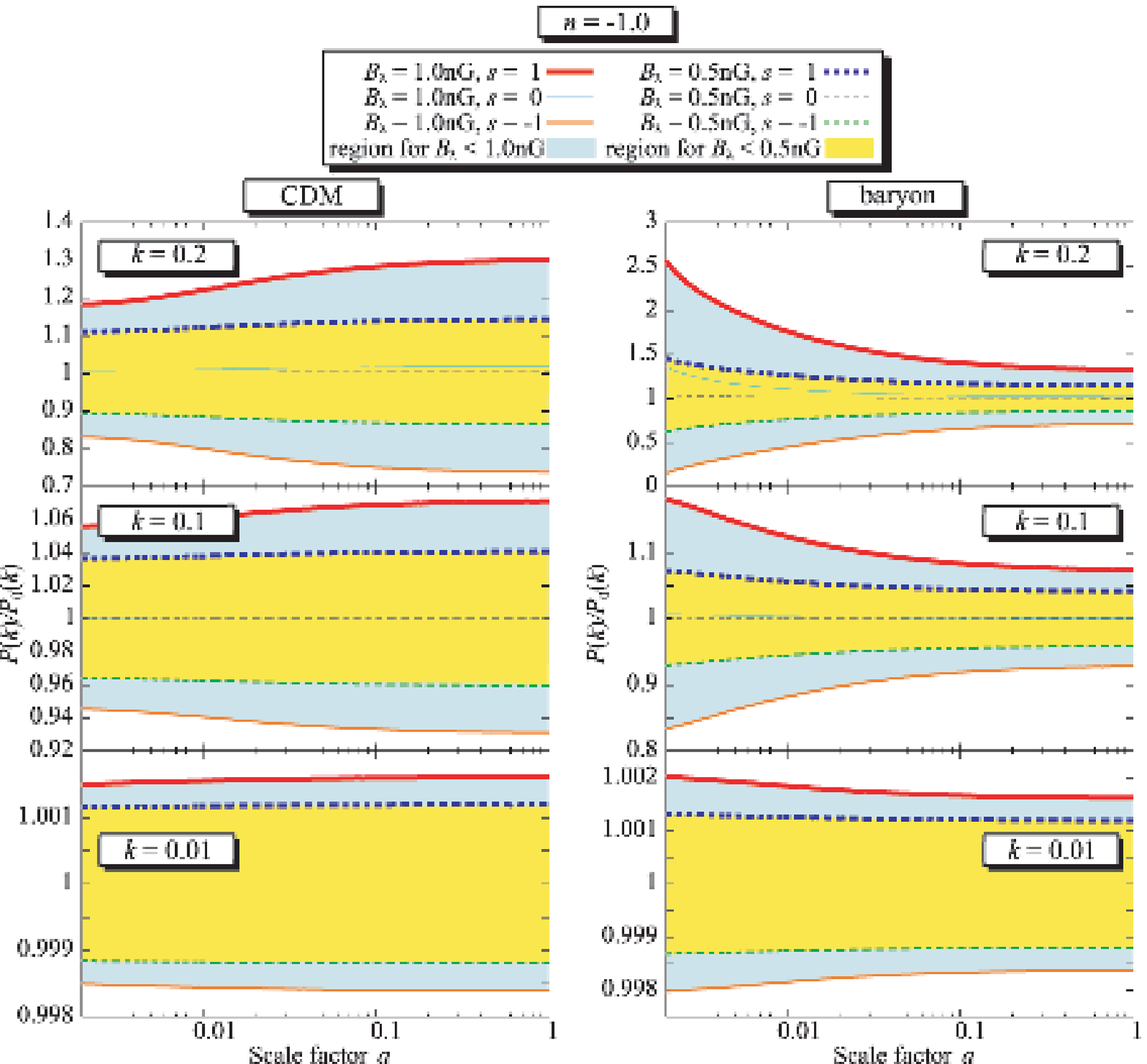}
\caption{\label{fig4}
Same as Fig.\ref{fig3}, but for a different power spectral index of the
 PMF $n=$ -1.0. 
}
\end{figure*}

\begin{figure*}[h]
\includegraphics{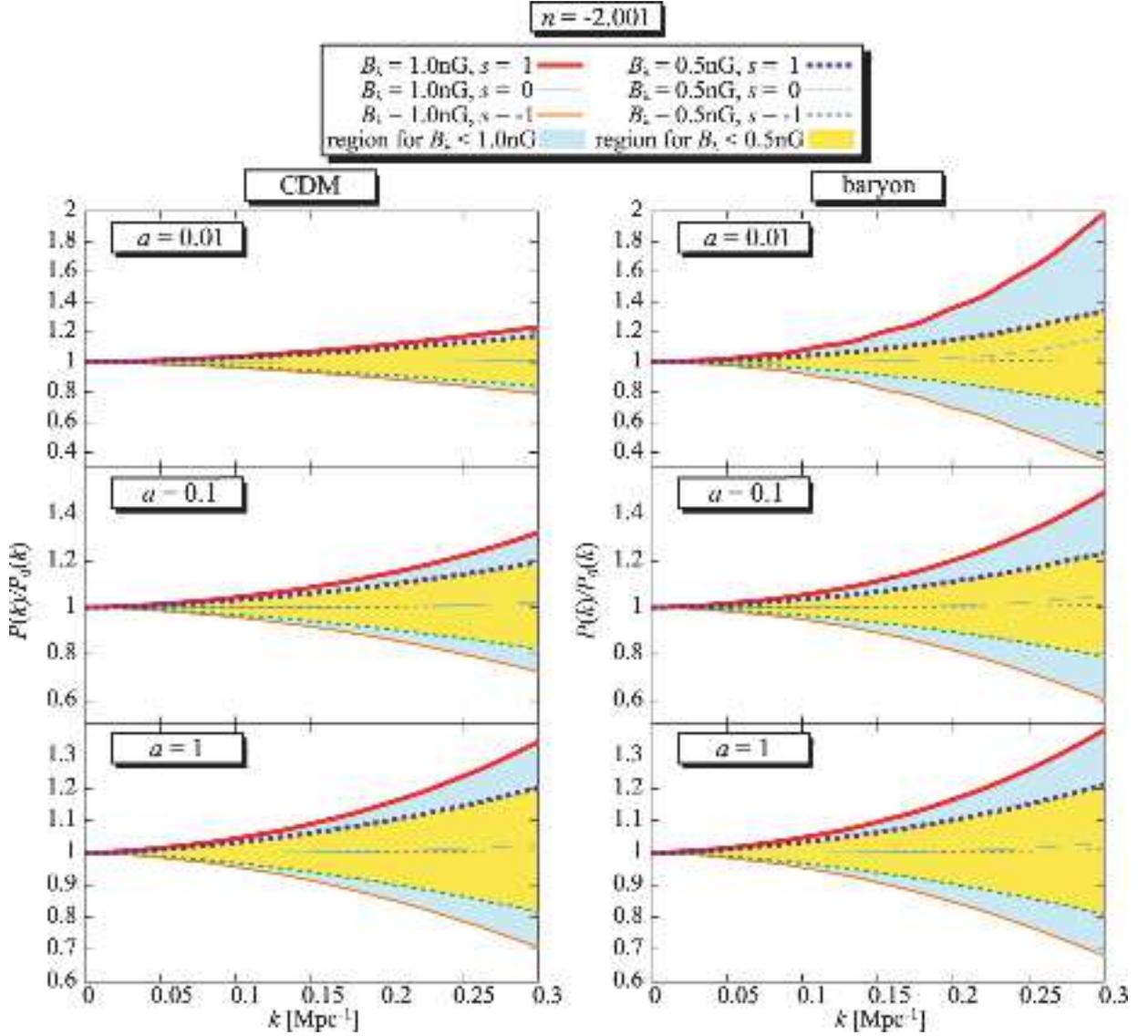}
\caption{\label{fig5}
Differences of transfer functions of CDM and baryon with ($P(k)$) and
 without ($P_0(k)$) the PMF, normalized by $P_0(k)$
  as a function of wave number $k$ at each redshift as indicated. 
The curves in right and left panels correspond to the differences in CDM
  and baryon, respectively.
The red thick, light-blue thin, orange normal, blue dotted bold, black dotted thin, and gree dotted middle curves in each both side panels show 
$(B_\lambda ,s)=$ 
(1.0nG,1),
(1.0nG,0),
(1.0nG,-1),
(0.5nG,1),
(0.5nG,0), and
(0.5nG,-1),
respectively.
In all figures, the power spectral index of the PMF is fixed to $n=$ -2.001.
The curves $s=0$ in all panels look like unit, acutually, those is more than unit.
}
\end{figure*}

\begin{figure*}[h]
\includegraphics{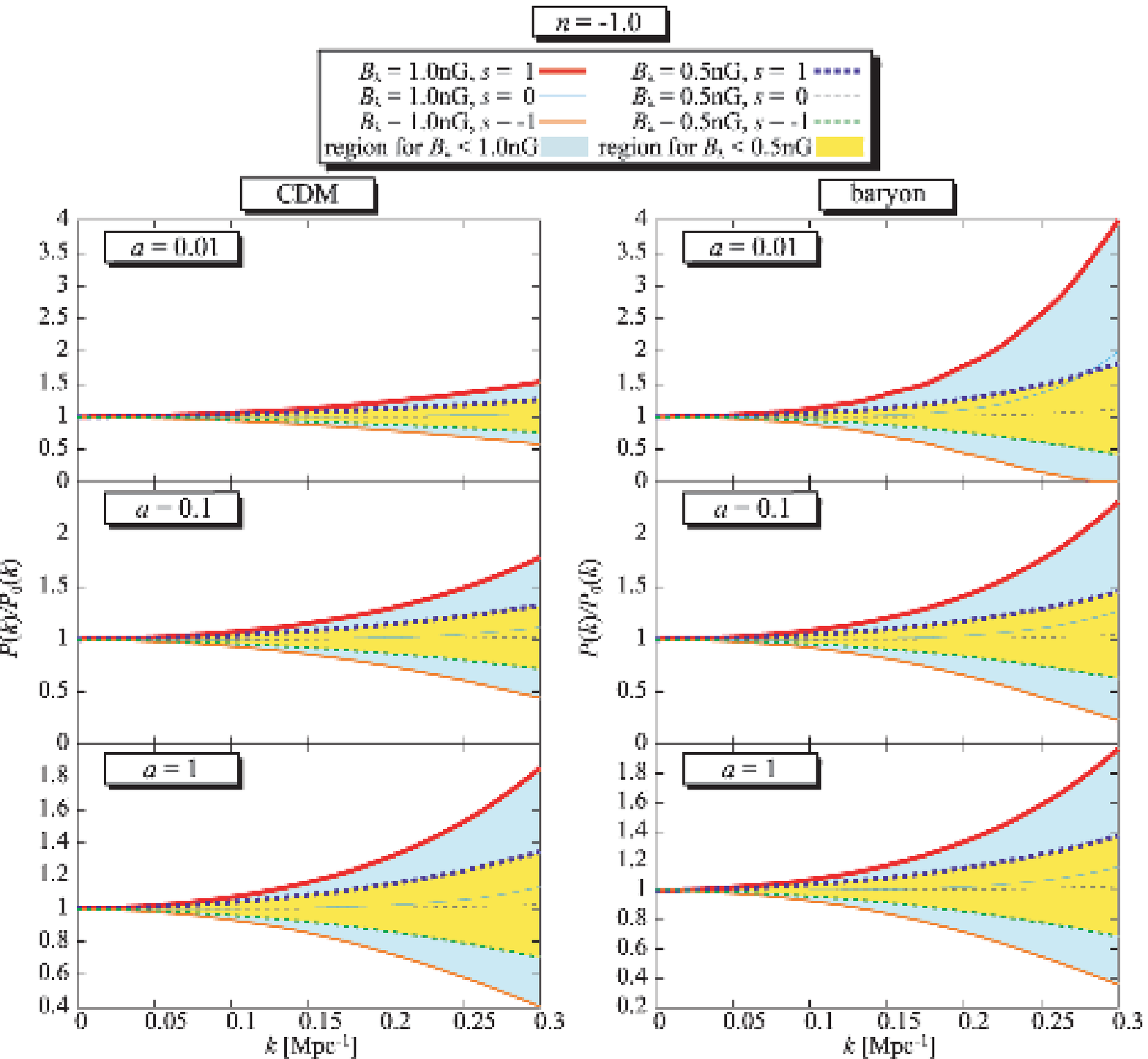}
\caption{\label{fig6}
Same as Fig.\ref{fig5}, 
 but for a different power spectral index of the PMF $n_\mathrm{B}=$ -1.0.}
\end{figure*}

After baryons decoupling from photons, 
the baryons density evolution starts to affect CDM density evolution
through gravitational interaction \cite{Yamamoto:1997qc}.
So the PMF effect on the CDM increases with time through baryons (left
panels in Figs. \ref{fig3} and \ref{fig4}).
Once the CDM density fluctuations are generated by the PMF indirectly, they
grow due to the gravitational instability so that the growthrate is the
same as that of the primordial CDM density fluctuations. 
The baryons density fluctuations are generated by PMF directly, 
and therefore, the gravitational potential 
does not dominate the evolution of baryon density fluctuation
 until $4\pi G \rho \, \delta \sim k^2 \frac{\Pi_{\mathrm{[EM:S]}}(\mbi{k},\tau)}{4\pi \rho_b}$.
The baryons density fluctuations with the PMF increase with the
wavenumber $k$ (right panels in Figs. \ref{fig5} and \ref{fig6}).
The reason is that the effect of the PMF is
relatively larger at smaller scales (larger $k$ values) in
Eq.(\ref{eq:baryon_v}). 
From Figs. \ref{fig5} and \ref{fig6} we found that 
the amplitude ratio between the density spectra with and without PMF 
($|P(k)/P_0(k)|$ at $k>0.2$ Mpc$^{-1}$)
lies between $75\%$ and $130\%$
at present for the range of PMF parameters $n =$ -2.001 and -1.0, 
$0.5$ nG $< B_\lambda < 1.0$ nG, 
and $-1<s<1$.
\section{\label{s:Discussions}Discussions of PMF model}
Taking into account the stochastic PMF power spectrum sets
(eqs.(\ref{ED_Souce})-(\ref{S_Souce})), we can investigate the most
accurate effect of PMF on the evolution of photons, baryons and CDM
density fluctuation.
From figs.\ref{fig3}-\ref{fig6}, it is turned out that there is a strong
degeneracy of the correlation factor $s_\mathrm{[DF]}$, and the PMF
amplitude $B_\lambda$. 
The correlation factor $s_\mathrm{[DF]}$ depends on the PMF origin.
So, in order to find a clue for solving such degeneracy problem, we
shall discuss a relation of the correlation factor $s_\mathrm{[DF]}$ and
the PMF origin. 
\subsection{\label{ssec:ne}Negative correlation of the density perturbations}
When $s_\mathrm{[DF]}<0$, the tension of PMF delays the evolution of the matter density fluctuation, and the pressure of PMF accelerates it.
In this case, therefore, when the PMF pressure for $n < -1.5$ dominates,
the evolution of the matter density fluctuation is accelerated by the PMF.
When $n > -1.5$, on the other hand,  the evolution is delayed by the PMF.
In order to substantiate such condition, 
the PMF must have been generated in the lower density regions.
In order to substantiate such condition, 
the PMF must have been generated in the lower density regions.
Some previous works mention that the PMF is directly proportional to the matter density on cosomological scale\cite{1,2}.
In these cases, since it is natural that there are more amplitude of PMF on the
higher density regions in the frozen-in, this condision would be
difficult to be realized with the causally generated PMF.
\subsection{\label{ssec:no}No correlation}
In the case of the no correlation between 
the PMF and the matter density fluctuations, 
the power spectrum function of the matter denasity fluctuation with the PMF 
is increased by the PMF, independently dominances of PMF pressure or tension.
Here we must notice the difference between the power spectrum function
$P(k)$ and the matter density fluctuation $\delta$. 
While the total density fluctuation $\delta$ can be smaller or larger
if the effect of PMF is dominated by its pressure or by its tension,
respectively, the power spectrum function $P(k)$ is always increased
when PMF does not correlate with primordial density fluctuations.
If the PMF is generated by density fluctuations proposed by
\cite{Ichiki:2006aa}, peaks of PMF are at peaks of the pressure
gradient of cosmic fluids. Consequently, the PMF is created along the
border between high and low density regions, i.e., $\delta\sim0$.
In this case there would be no (or very weak) correlation between the
PMF and density perturbation statistically.

\subsection{\label{ssec:p}Positive correlation}
When $s_\mathrm{[DF]}>0$, the pressure of PMF delays the evolution of the matter density fluctuation, and the tension of PMF accelerates it.
So, in this case, when the PMF pressure dominates for $n < -1.5$ ,
the evolution of the matter density fluctuation is delayed by the PMF. 
In order to substantiate such condition, the PMF may be generated on the
higher energy density regions. 
As an example, let us consider the PMF which was generated by a vector
potential generated from the dilaton during inflation\cite{bamba04}.
If the coupling between fields of dilaton and inflaton is negligibly
small, there would be no correlation between the PMF and density
perturbations. This is because the PMF was generated by the vector field
coupled with dilaton, while the inflaton was responsible for density
perturbations.
However, if we consider the case that a curvature coupling (like $R
F_{\mu\nu}F^{\mu\nu}$) generates the PMF in the same time, the positive
correlation between PMF and density fields would be expected, since the 
electromagnetic fields are coupled with $R$ or Hubble
parameter, which is determined by the density fields.
Because, as mentioned above Subsection.\ref{ssec:ne}, the positive correlation between the magnetic field and the matter density at present is reasonable\cite{1,2} without any surprising PMF generations after the inflation, the positive correlation is the most natural consequence of such (inflationary) PMF generation models.
\\
\section{\label{s:con}Summary}
We numerically investigated the effect of the PMF on the energy density
fields by considering the stochastic one that depends on scales, and we
quantitatively discuss the effect of the PMF on the seeds of LSS in the
early universe.
We considered more general effects of the PMF than those considered in
the previous works.  We considered not only the magnetic field 
tension but also the increases of pressure and energy
density perturbations from the field.
Furthermore, by considering the correlation between the PMF and the
matter density fluctuation, and taking the mathematically exact
stochastic PMF power spectrum sets,  we obtained reasonable and accurate
evolutions of 
baryon, CDM, photon, and therefore the large scale structure. 
We show that the PMF can play very different roles on the evolution of
density perturbations accoding to the correlation. 
After decoupling,
CDM is also influenced indirectly by the PMF through
gravitational interaction.  
We have estimated the effects, and we found that 
the amplitude ratio between the density spectra with and without PMF 
($|P(k)/P_0(k)|$ at $k>0.2$ Mpc$^{-1}$)
lies between $75\%$ and $130\%$
at present for the range of PMF parameters $n =$ -2.001 and -1.0, 
$0.5$ nG $< B_\lambda < 1.0$ nG, 
and $-1<s<1$.

Interestingly, it is reported that the magnetic field at large scales
($\lambda=$1Mpc) around $B_\lambda \sim $ nG \cite{Mack02,Lewis04,yamazaki05b,yamazaki06,Durrer:1999bk} provides a
new interpretation for the excess of CMB anisotropies at smaller angular
scales. If the PMF with such strength was present, it is very likely
that it has affected the formations at large scale structure as shown in
the present paper.
Yoshida, Sugiyama and Hernquist \cite{Yoshida:2003sy} suggested that in
order to avoid false coupling of the baryon and CDM for small scales,
using independent transfer functions for the baryon and CDM is
preferable. The PMF would be another source of this
difference in the transfer function for baryon and CDM. 
Since the density perturbations in the early universe have evolved to
the LSS at present age, the evolution of the LSS with the PMF becomes
more different than that without the PMF. We have shown that
the baryon and CDM energy density perturbations follow very different
evolutions in the presence of the PMF;  with the PMF taken into
consideration, the evolution of large scale structure should become more
complicated.

\begin{acknowledgments}
We thank K. Omukai for a critical comment on the early version of our
 paper, and K. Bamba for discussion on the PMF generation during inflation.
We are also grateful to S. Kawagoe, M. Kusakabe, S. Inoue, G. J. Mathews and
 T. Kajino for their valuable discussions. 
D.G.Y. and K. I. acknowledge the support by Grant-in-Aid for JSPS Fellows.
\end{acknowledgments}
\newpage 
\appendix
\section{Power Spectrum of PMF}
In this section we derive power spectral of PMF, $Z$, $\Pi$ and $E$,
which we used in sec.\ref{sec:eff}.
The electro-magnetic energy momentum tensor can be decomposed into three
parts, i.e.,  scalar, vector, and tensor parts.
The scalar part of electro-magnetic energy momentum tensor
$T_{ij}{}_{[\mathrm{EM}:S]}$ is definied as, 
\begin{eqnarray}
T_{ij}{}_{[\mathrm{EM}:S]} 
	= ^{S}\mathcal{P}_{ijlm}T^{lm}{}_{[\mathrm{EM}]},\label{tran}
\end{eqnarray}
where $^{S}\mathcal{P}_{ijlm}$ is a scalar project tensor:
\begin{eqnarray}
^{S}\mathcal{P}_{ijlm}=
       \hat{k}_j(\hat{k}_mP_{il} - \hat{k}_lP_{im})
 +     \frac{1}{2}P_{ij}P_{lm}
 +     \hat{k}_i\hat{k}_j\hat{k}_l\hat{k}_m ~,\label{ap1}
\end{eqnarray}
and
\begin{eqnarray}
P^{ij}(k')&=&
	\delta^{ij}-\frac{k'{}^{i}k'{}^{j}}{k'{}^2}.\label{project_tensor}
\end{eqnarray}
Therefore it is easily shown that
\begin{eqnarray}
\hat{k}^i\hat{k}^j{}^{S}\mathcal{P}_{ijlm}=\hat{k}_l\hat{k}_m. \label{ap1.1}
\end{eqnarray}
\subsection{Power spectrum of Lorenz Force: $\Pi(k)$}
Using Eqs. (\ref{ap1})-(\ref{ap1.1}), 
we obtain a two-point correlation function of scalar part
\begin{eqnarray}
\langle 
	T_{[\mathrm{EM:S}]}(\mbi{k})
	T^*_{[\mathrm{EM:S}]}(\mbi{k}')
\rangle
&=& \hat{k}_i\hat{k}^{j}\hat{k}_l\hat{k}^m
\langle
 	T^{i}_{j}{}_{[\mathrm{EM}]}(\mbi{k})
 	T^{*l}_{m}{}_{[\mathrm{EM}]}(\mbi{k}')
\rangle\nonumber\\
&=&
(2\pi)^3|\Pi_{[\mathrm{EM:S}]}(\mbi{k})|^2\delta(\mbi{k}-\mbi{k}').\label{eq:TPCFofLF}
\end{eqnarray}
where we define a power spectrum of Lorenz force
$\Pi_{[\mathrm{EM:S}]}(\mbi{k})$. 
The electromagnetic stress-energy tonsor in $k$ space is given by 
\begin{eqnarray}
T^{i}_{j}(\mbi{k})_{[\mathrm{EM}]}
 &=& \frac{1}{4\pi a^{4}}\int\frac{d^3k'}{(2\pi)^3}
 	\left\{ 
 		\frac{1}{2}\delta^{i}_{j}B^{l}(\mbi{k}') B_{l}(\mbi{k}-\mbi{k}') 
 		-\frac{}{}B^{i}(\mbi{k}') B_{j}(\mbi{k}-\mbi{k}')
 	\right\}. \label{ap2} 
\end{eqnarray}
For convenience, we decompose $T^{i}_{j}(\mbi{k})_{[\mathrm{EM}]}$ into two
parts as follows,
\begin{eqnarray}
T^{i}_{j}(\mbi{k})_{[\mathrm{EM}]}
	&=&T^{i}_{j}(\mbi{k})_{[\mathrm{EM:1}]}
	-T^{i}_{j}(\mbi{k})_{[\mathrm{EM:2}]}
	\label{ap3}\\
T^{i}_{j}(\mbi{k})_{[\mathrm{EM:}1]}
	&=& \frac{1}{4\pi a^{4}}\int\frac{d^3k'}{(2\pi)^3}
 	\frac{1}{2}\delta^{i}_{j}B^{l}(\mbi{k}') B_{l}(\mbi{k}-\mbi{k}') \label{ap4} \\
T^{i}_{j}(\mbi{k})_{[\mathrm{EM:}2]}
 	&=& \frac{1}{4\pi a^{4}}\int\frac{d^3k'}{(2\pi)^3}
B^{i}(\mbi{k}') B_{j}(\mbi{k}-\mbi{k}') \label{ap5}.
\end{eqnarray}
Correspondingly, we define 
$T_\mathrm{[EM:S1]}=\hat{k}_i\hat{k}^j T^{i}_{j}{}_\mathrm{[EM:1]}$ and 
$T_\mathrm{[EM:S2]}=\hat{k}_i\hat{k}^j T^{i}_{j}{}_\mathrm{[EM:2]}$. 
Using Eq.(\ref{ap3} - \ref{ap5}), we can rewrite the two point correlation function of Lorenz force of scalar part as,
\begin{eqnarray}
\langle
T(\mbi{k})_\mathrm{[EM:S]}T^*(\mbi{k}')_\mathrm{[EM:S]}
\rangle
&=&
\langle
(T(\mbi{k})_\mathrm{[EM:S1]}-T(\mbi{k})_\mathrm{[EM:S2]})
(T^*(\mbi{k}')_\mathrm{[EM:S1]}-T^*(\mbi{k}')_\mathrm{[EM:S2]})
\rangle  
\nonumber\\
&=&
\langle
T(\mbi{k})_\mathrm{[EM:S1]}T^*(\mbi{k}')_\mathrm{[EM:S1]}
\rangle  
-\langle
T(\mbi{k})_\mathrm{[EM:S1]}T^*(\mbi{k}')_\mathrm{[EM:S2]}
\rangle  
\nonumber\\
&-&
\langle
T(\mbi{k})_\mathrm{[EM:S2]}T^*(\mbi{k}')_\mathrm{[EM:S1]}
\rangle  
+\langle
T(\mbi{k})_\mathrm{[EM:S2]}T^*(\mbi{k}')_\mathrm{[EM:S2]}
\rangle. \label{ap6}  
\end{eqnarray}
Also, from (\ref{eq:TPCFofLF}), $\Pi_{[\mathrm{EM:S}]}(\mbi{k})$
in Eq.(\ref{eq:scalar_motion}) and (\ref{eq:baryon_v})
 can be written by,
\begin{eqnarray}
\Pi_{\mathrm{[EM:S]}}(\mbi{k})\delta(\mbi{k}-\mbi{k}')
=
\sqrt{
\frac{1}{(2\pi)^3}
\langle
T(\mbi{k})_{\mathrm{[EM:S]}}T^*(\mbi{k}')_{\mathrm{[EM:S]}}
\rangle
}.\label{LF_Souce}
\end{eqnarray}
\subsection{
Power spectrum of Shear Stress: $Z(\mbi{k})$
}
A power spectrum of Loreze force for the scalar part is decomposed into 
trace-trace and traceless (a shear pressure of PMF) parts.
A traceless part of electromagnetic stress-energy tensor can be written by,
\begin{eqnarray}
&&
\Sigma^i_j{}_\mathrm{[EM]}=
T^{i}_{j}{}_\mathrm{[EM]}-\frac{1}{3}\delta^{i}_{j}T^{k}_{k}{}_\mathrm{[EM]}
\nonumber\\
&&	= 
	\frac{1}{4\pi a^{4}}\int\frac{d^3k'}{(2\pi)^3}
 	\left\{
 		\frac{1}{3}\delta^{i}_{j}B^{l}(k') B_{l}(k-k')
		-\frac{}{}B^{i}(k') B_{j}(k-k')
	\right\}. \nonumber\\
	\label{ap7}
\end{eqnarray}
Thus, a scalar part of traceless component is given by,
\begin{eqnarray}
&&
\Sigma_\mathrm{[EM:S]}(\mbi{k})=
	\left(
		\hat{k}_i\hat{k}^j-\frac{1}{3}\delta_{i}^{j}
	\right)
	\Sigma^i_j{}_\mathrm{[EM]}
\nonumber\\
&&	= 
	\frac{1}{4\pi a^{4}}\int\frac{d^3k'}{(2\pi)^3}
	\hat{k}_i\hat{k}^j
 	\left\{
 		\frac{1}{3}\delta^{ij}B^{l}(k') B_{l}(k-k')
		-B^{i}(k') B_{j}(k-k')
	\right\}
\nonumber\\
&&	= 
 		\left\{
 			\frac{2}{3}T(\mbi{k})_\mathrm{[EM:S1]}
 			-T(\mbi{k})_\mathrm{[EM:S2]}
 		\right\}
\end{eqnarray}
Similarly as the above subsection, we obtain the two point correlation function of PMF shear stress for scalar part as the following,
\begin{eqnarray}
&&\langle
	\Sigma(\mbi{k})_\mathrm{[EM:S]}
	\Sigma(\mbi{k}')^*_\mathrm{[EM:S]}
\rangle
\nonumber\\
&&=
\frac{4}{9}
\langle
	T(\mbi{k})_\mathrm{[EM:S1]}T^*(\mbi{k}')_\mathrm{[EM:S1]}
\rangle  
-\frac{2}{3}\langle
	T(\mbi{k})_\mathrm{[EM:S1]}T^*(\mbi{k}')_\mathrm{[EM:S2]}
\rangle  
\nonumber\\
&&
\ \ \ \ \ \ -\frac{2}{3}\langle
	T(\mbi{k})_\mathrm{[EM:S2]}T^*(\mbi{k})_\mathrm{[EM:S1]}
\rangle  
+\langle
	T(\mbi{k})_\mathrm{[EM:S2]}T^*(\mbi{k})_\mathrm{[EM:S2]}
\rangle  
\nonumber\\
&&=(2\pi)^3|Z(\mbi{k})_{\mathrm{[EM:S]}}|^2\delta(\mbi{k}-\mbi{k}'),
\label{PMF_Shear}
\end{eqnarray}
where $Z(\mbi{k})_{\mathrm{[EM:S]}}$ is a power spectrum of PMF shear stress.
We rewrite Eq.(\ref{PMF_Shear}) as following,
\begin{eqnarray}
Z^2(\mbi{k})_{\mathrm{[EM:S]}}\delta(\mbi{k}-\mbi{k}')
=
\frac{1}{(2\pi)^3}
\langle
	\Sigma(\mbi{k})_\mathrm{[EM:S]}
	\Sigma(\mbi{k}')^*_\mathrm{[EM:S]}
\rangle
.\label{S_Soucea}
\end{eqnarray}
\subsection{
Power spectrum of PMF energy:$E(\mbi{k})$
}
The PMF energy is defined by Eq.(\ref{eq_MST_00}).
The PMF energy in $k$ space is given by 
\begin{eqnarray}
T^{0}_{0}(\mbi{k})_{[\mathrm{EM}]}
 = \frac{1}{8\pi a^{4}}\int\frac{d^3k'}{(2\pi)^3}
 		B^{l}(\mbi{k}') B_{l}(\mbi{k}-\mbi{k}')
 =T(\mbi{k})_{[\mathrm{EM:S1}]}.
 	 \label{ps_pmf_e} 
\end{eqnarray}
Similarly as the above subsection, we obtain the two point correlation function of PMF energy for scalar part as the following,
\begin{eqnarray}
\langle
	T(\mbi{k})_\mathrm{[EM:S1]}
	T(\mbi{k}')^*_\mathrm{[EM:S1]}
\rangle
=(2\pi)^3|E(\mbi{k})_{\mathrm{[EM:S]}}|^2\delta(\mbi{k}-\mbi{k}'),
\label{PMF_Energy}
\end{eqnarray}
where $E(\mbi{k})_{\mathrm{[EM:S]}}$ is a power spectrum of PMF energy.
We rewrite Eq.(\ref{PMF_Energy}) as following,
\begin{eqnarray}
E^2(\mbi{k})_{\mathrm{[EM:S]}}\delta(\mbi{k}-\mbi{k}')
=
\frac{1}{(2\pi)^3}
\langle
	T(\mbi{k})_\mathrm{[EM:S1]}
	T(\mbi{k}')^*_\mathrm{[EM:S1]}
\rangle
.\label{E_Soucea}
\end{eqnarray}

In order to obtain 
$\Pi(\mbi{k})_{\mathrm{[EM:S]}}$, 
$Z(\mbi{k})_{\mathrm{[EM:S]}}$, and 
$E(\mbi{k})_{\mathrm{[EM:S]}}$, 
we calculate 
$
\langle
	T_{[\mathrm{EM:S1}]}(\mbi{k})
	T^*_{[\mathrm{EM:S1}]}(\mbi{k})
\rangle
$
,
$
\langle
	T_{[\mathrm{EM:S1}]}(\mbi{k})
	T^*_{[\mathrm{EM:S2}]}(\mbi{k})
\rangle
	+
\langle
	T_{[\mathrm{EM:S2}]}(\mbi{k})
	T^*_{[\mathrm{EM:S1}]}(\mbi{k})
\rangle
$
, and
$
\langle
	T_{[\mathrm{EM:S2}]}(\mbi{k})
	T^*_{[\mathrm{EM:S2}]}(\mbi{k})
\rangle
$
in Eqs.(\ref{ap6}), (\ref{PMF_Shear}) and (\ref{PMF_Energy}) the following subsections. 
\subsection{
$
\langle
	T_{[\mathrm{EM:S1}]}(\mbi{k})
	T^*_{[\mathrm{EM:S1}]}(\mbi{k})
\rangle
$}
Using Eq.(\ref{tran}), (\ref{ap1}) and (\ref{ap4}), 
the scalar part of $T(\mbi{k})_{[\mathrm{EM:}1]}$ becomes
\begin{eqnarray}
T(\mbi{k})_{[\mathrm{EM:S}1]}
	&=&	\hat{k}_i\hat{k}^j T^{i}_{j}(\mbi{k})_{[EM:1]}\nonumber\\
	&=& \hat{k}_i\hat{k}^j \frac{1}{8\pi a^{4}}
		\int\frac{d^3k'}{(2\pi)^3}
 	 	\delta^{i}_{j}B^{l}(\mbi{k}') B_{l}(\mbi{k}-\mbi{k}') \nonumber \\
	&=& \frac{1}{2^2(2\pi)^4 a^{4}}
		\int d^3k' B^{l}(\mbi{k}') B_{l}(\mbi{k}-\mbi{k}'). \nonumber \\
\label{eq:T_EM_S1_1} 
\end{eqnarray}
Thus the two point correlation function of Eq.(\ref{eq:T_EM_S1_1}) given by 
\begin{eqnarray}
&&\langle 
	T(\mbi{k})_{[\mathrm{EM:S1}]}
	T(\mbi{p})^*_{[\mathrm{EM:S1}]}
\rangle
	= \nonumber\\
&&	\frac{1}{2^4(2\pi)^8 a^{8}}\int d^3k'd^3p'
	  \langle B^{i}(\mbi{k}') B_{i}(\mbi{k}-\mbi{k}')
 	  B^{l*}(\mbi{p}') B^*_{l}(\mbi{p}-\mbi{p}')\rangle\nonumber \\
\label{eq:T_EM_S1_2}
\end{eqnarray}
We assume the random magnetic field is Gaussian and apply the Wick's theorem
\begin{eqnarray}
\left\langle B^{i}(\mbi{k}') B_{i}(\mbi{k}-\mbi{k}'){B^{l}}^*(\mbi{p}'){B_{l}}^*(\mbi{p}-\mbi{p}')\right\rangle 
&=&		\left\langle B^{i}(\mbi{k}') B_{i}(\mbi{k}-\mbi{k}')\right\rangle
		\left\langle {B^{l}}^*(\mbi{p}'){B_{l}}^*(\mbi{p}-\mbi{p}')\right\rangle \nonumber \\
	&&+	\left\langle B^{i}(\mbi{k}'){B^{l}}^*(\mbi{p}')\right\rangle
		\left\langle B_{i}(\mbi{k}-\mbi{k}'){B_{l}}^*(\mbi{p}-\mbi{p}')\right\rangle\nonumber \\
	&&+	\left\langle B^{i}(\mbi{k}'){B_{l}}^*(\mbi{p}-\mbi{p}')\right\rangle
		\left\langle B_{i}(\mbi{k}-\mbi{k}'){B^{l}}^*(\mbi{p}')\right\rangle .\nonumber \\
		\label{eq:InMPforSangle}
\end{eqnarray}
A two-point correlation funciton for PMF is defined by Ref.\cite{Mack02}
\begin{eqnarray}
\left\langle B^{i}(\mbi{k}) {B^{j}}^*(\mbi{k}')\right\rangle 
	&=&	\frac{(2\pi)^{n+8}}{2k_\lambda^{n+3}}
		\frac{B^2_{\lambda}}{\Gamma\left(\frac{n+3}{2}\right)}
		k^nP^{ij}(k)\delta(\mbi{k}-\mbi{k}').
		\label{two_point1} 
\end{eqnarray}
Using the reality condition
\begin{eqnarray}
B^{i}(\mbi{k})={B^{i}}^*(-\mbi{k}),
\end{eqnarray}
and substituting (\ref{two_point1}) into (\ref{eq:InMPforSangle}),
first, second, and thrid terms become 
\begin{eqnarray}
\left\langle B^{i}(\mbi{k}') B_{i}(\mbi{k}-\mbi{k}')\right\rangle 
\left\langle {B^{l}}^*(\mbi{p}') {B_l}^*(\mbi{p}-\mbi{p}')\right\rangle 
	=	\left\{
		\frac{(2\pi)^{n+8}}{2k_\lambda^{n+3}}
		\frac{B^2_{\lambda}}{\Gamma\left(\frac{n+3}{2}\right)}
	\right\}^2
	k'{}^n(-p')^n
	4\delta(\mbi{k})\delta(-\mbi{p}),\nonumber\\ \label{eq:MPforS01}
\end{eqnarray}
\begin{eqnarray}
\left\langle {B^{i}}(\mbi{k}') {B^{l}}^*(\mbi{p}')\right\rangle 
\left\langle {B_{i}}(\mbi{k}-\mbi{k}') {B_{l}}^*(\mbi{p}-\mbi{p}')\right\rangle &=&	\left\{
		\frac{(2\pi)^{n+8}}{2k_\lambda^{n+3}}
		\frac{B^2_{\lambda}}{\Gamma\left(\frac{n+3}{2}\right)}
	\right\}^2
	k'{}^n|\mbi{k}-\mbi{k}'|^n
	P^{il}(k')P_{il}(k-k')
\nonumber\\
&&\times
	\delta(\mbi{k}'-\mbi{p}')
	\delta((\mbi{k}-\mbi{k}')-(\mbi{p}-\mbi{p}')),
	\nonumber\\ \label{eq:MPforS02}
\end{eqnarray}
and 
\begin{eqnarray}
\left\langle {B^{i}}(\mbi{k}') {B_{l}}^*(\mbi{p}-\mbi{p}')\right\rangle 
\left\langle {B_{i}}(\mbi{k}-\mbi{k}') {B^{l}}^*(\mbi{p}')\right\rangle 
&=&	\left\{
		\frac{(2\pi)^{n+8}}{2k_\lambda^{n+3}}
		\frac{B^2_{\lambda}}{\Gamma\left(\frac{n+3}{2}\right)}
	\right\}^2
	k'{}^n|\mbi{k}-\mbi{k}'|^n
	P^i_l(k')P^l_i(k-k')
\nonumber\\
&&\times
	\delta((\mbi{k}-\mbi{k}')-\mbi{p}')
	\delta(\mbi{k}'-(\mbi{p}-\mbi{p}')),
\nonumber\\ \label{eq:MPforS05}
\end{eqnarray}
respectively.
Because $\mbi{k}\ne 0$($x=2\pi/k\ne\infty$), 
$\delta(\mbi{k})=\delta(-\mbi{p})=0$,
the first term on the r.h.s.in Eq.(\ref{eq:InMPforSangle}) is 0.
From Eq.(\ref{project_tensor})
\begin{eqnarray}
P^{il}(k')P_{il}(k-k')
=
	1+\frac{\{\mbi{k}'\cdot(\mbi{k}-\mbi{k}')\}^2}{k'{}^2(k-k')^2}.
	 \label{project_tensor2}
\end{eqnarray}
Therefore second and third terms in Eq.(\ref{eq:InMPforSangle}) become 
\begin{eqnarray}
&&
	\left\langle
		{B^{i}}(\mbi{k}') {B^{l}}^*(\mbi{p}')
	\right\rangle 
	\left\langle
		{B_{i}}(\mbi{k}-\mbi{k}') {B_{l}}^*(\mbi{p}-\mbi{p}')
	\right\rangle \nonumber\\ 
&&=
	\left\{
		\frac{(2\pi)^{n+8}}{2k_\lambda^{n+3}}
		\frac{B^2_{\lambda}}{\Gamma\left(\frac{n+3}{2}\right)}
	\right\}^2
	k'{}^n|\mbi{k}-\mbi{k}'|^n
	\left\{
		1+\frac{\{\mbi{k}'\cdot(\mbi{k}-\mbi{k}')\}^2}
			   {k'{}^2|\mbi{k}-\mbi{k}'|^2}
	\right\}
	\delta(\mbi{k}'-\mbi{p}')
	\delta((\mbi{k}-\mbi{k}')-(\mbi{p}-\mbi{p}'))\nonumber\\ 
&& \label{secondterm}
\end{eqnarray}
and
\begin{eqnarray}
&&
\left\langle
	{B^{i}}(\mbi{k}') {B_{l}}^*(\mbi{p}-\mbi{p}')
\right\rangle
\left\langle
	{B_{i}}(\mbi{k}-\mbi{k}') {B^{l}}^*(\mbi{p}')
\right\rangle \nonumber\\ 
&&=
	\left\{
		\frac{(2\pi)^{n+8}}{2k_\lambda^{n+3}}
		\frac{B^2_{\lambda}}{\Gamma\left(\frac{n+3}{2}\right)}
	\right\}^2
	k'{}^n|\mbi{k}-\mbi{k}'|^n
	\left\{
		1+\frac{\{\mbi{k}'\cdot(\mbi{k}-\mbi{k}')\}^2}
			   {k'{}^2|\mbi{k}-\mbi{k}'|^2}
	\right\}
\delta(\mbi{k}'-(\mbi{p}-\mbi{p}'))
\delta((\mbi{k}-\mbi{k}')-\mbi{p}'),
\label{thirdterm}
\end{eqnarray}
respectively.
Using Eq. (\ref{secondterm}) and (\ref{thirdterm}), 
Eq.(\ref{eq:T_EM_S1_2}) becomes
\begin{eqnarray}
&&	\langle
		T(\mbi{k},\tau)_{[\mathrm{EM:S1}]}
		T^*(\mbi{p},\tau)_{[\mathrm{EM:S1}]}
	\rangle
\nonumber \\
&&=
	\frac{1}{2^4(2\pi)^8 a^{8}}\int d^3k'd^3p'
	\left\{
		\frac{(2\pi)^{n+8}}{2k_\lambda^{n+3}}
		\frac{B^2_{\lambda}}{\Gamma\left(\frac{n+3}{2}\right)}
	\right\}^2
	k'{}^n|\mbi{k}-\mbi{k}'|^n
	\left\{
		1+\frac{\{\mbi{k}'\cdot(\mbi{k}-\mbi{k}')\}^2}
			   {k'{}^2|\mbi{k}-\mbi{k}'|^2}
	\right\}\nonumber\\
&&\times
	\{
		\delta(\mbi{k}'-\mbi{p}')\delta((\mbi{k}-\mbi{k}')-(\mbi{p}-\mbi{p}'))
		+\delta(\mbi{k}'-(\mbi{p}-\mbi{p}'))\delta((\mbi{k}-\mbi{k}')-\mbi{p}')
	\}
\nonumber\\
\label{eq:T_EM_S1_2}
\end{eqnarray}
Integrating Eq.(\ref{eq:T_EM_S1_2}) by $p'$,
we obtain following equation
\begin{eqnarray}
&&	\langle
		T(\mbi{k},\tau)_{[\mathrm{EM:S1}]}
		T(\mbi{p},\tau)^*_{[\mathrm{EM:S1}]}
	\rangle
\nonumber \\
&&=
	\frac{2}{2^4(2\pi)^8 a^{8}}\int d^3k'
	\left\{
		\frac{(2\pi)^{n+8}}{2k_\lambda^{n+3}}
		\frac{B^2_{\lambda}}{\Gamma\left(\frac{n+3}{2}\right)}
	\right\}^2
	k'{}^n|\mbi{k}-\mbi{k}'|^n
	\left\{
		1+\frac{\{\mbi{k}'\cdot(\mbi{k}-\mbi{k}')\}^2}
			   {k'{}^2|\mbi{k}-\mbi{k}'|^2}
	\right\}
	\delta(\mbi{k}-\mbi{p}).
\nonumber\\
\label{eq:T_EM_S1_3}
\end{eqnarray}
We define that
\begin{eqnarray}
	\mathcal{C}
		= \cos{c}
		= \hat{\mbi{k}}\cdot\hat{\mbi{k}}'
		= \frac{\mbi{k}'\cdot\mbi{k}}{k'k}.\label{define_C}
\end{eqnarray}
Substituting Eq.(\ref{define_C}) into Eq.(\ref{eq:T_EM_S1_3}), 
The two point function of Lorenz Force becomes
\begin{eqnarray}
	\langle
		T(\mbi{k},\tau)_{[\mathrm{EM:S1}]}
		T(\mbi{p},\tau)^*_{[\mathrm{EM:S1}]}
	\rangle
	&=&
	\frac{1}{2^3(2\pi)^8 a^{8}}
	\left\{
		\frac{(2\pi)^{n+8}}{2k_\lambda^{n+3}}
		\frac{B^2_{\lambda}}{\Gamma\left(\frac{n+3}{2}\right)}
	\right\}^2
\nonumber \\
&&\times
	\int d^3k'
		k'{}^n|\mbi{k}-\mbi{k}'|^{n-2}
		\left\{
			(1+\mathcal{C}^2)k^2-4kk'\mathcal{C}+2k'{}^2
		\right\}
	\delta(\mbi{k}-\mbi{p}).
\label{eq:T_EM_S1_4}
\end{eqnarray}
Choosing $\hat{\mbi{k}}$ to the polar axis as 
\begin{eqnarray}
d^3k'=k'{}^2dk'\sin{c}\ dc\ d\phi,
\end{eqnarray}
and integrating Eq.(\ref{eq:T_EM_S1_4}) by $\phi$,
the two point function of Lorenz Force is given by
\begin{eqnarray}
	\langle
		T(\mbi{k},\tau)_{[\mathrm{EM:S1}]}
		T(\mbi{p},\tau)^*_{[\mathrm{EM:S1}]}
	\rangle
	&=&
\frac{1}{2^3(2\pi)^7 a^{8}}
\left\{\frac{(2\pi)^{n+8}}{2k_\lambda^{n+3}}\frac{B^2_{\lambda}}{\Gamma\left(\frac{n+3}{2}\right)}\right\}^2\nonumber\\
&&\times
\int dk'k'{}^{n+2}\int^{1}_{-1} d\mathcal{C}
|\mbi{k}-\mbi{k}'|^{n-2}
\left\{
(1+\mathcal{C}^2)k^2-4kk'\mathcal{C}+2k'{}^2
\right\}
\delta(\mbi{k}-\mbi{p})
\nonumber\\  
&& \label{eq:T_EM_S1_5}
\end{eqnarray}
Almost all of the previous works have treated terms which include
$\mathcal{C}$ in the middle parenthesis as unity. 
In this paper, however, we calculate Eq.(\ref{eq:T_EM_S1_5}) furthur by
integrating by parts. 
In addition they have calculated integration of $k'$ using the Taylor expansion by $k'/k (k'\ll k)$ or $k/k' (k\ll k')$.
If we would want to estimate Eq.(\ref{eq:T_EM_S1_5}) for only $k'\ll k$
or $k\ll k'$, such approximation would be useful. 
However there is the value $k'\sim k$ in the integration range,
so we must caluculate Eq.(\ref{eq:T_EM_S1_5}) without such Taylor expansion.
Integrating Eq.(\ref{eq:T_EM_S1_4}) by $\mathcal{C}$,
after long but straightforward calculation, we can obtain following expression,
\begin{eqnarray}
	\langle
		T(\mbi{k},\tau)_{[\mathrm{EM:S1}]}
		T(\mbi{p},\tau)^*_{[\mathrm{EM:S1}]}
	\rangle
	&=&
	\frac{1}{4(2\pi)^7 a^{8}}
	\left\{
		\frac{(2\pi)^{n+8}}{2k_\lambda^{n+3}}
		\frac{B^2_{\lambda}}{\Gamma\left(\frac{n+3}{2}\right)}
	\right\}^2
\nonumber\\
&&\times	\int dk'k'{}^{n+2}
	\left[
	\frac{n^2+4n+1}{kk'n(n+2)(n+4)}
	\left\{
		(k+k')^{n+2}
		-|k-k'|^{n+2}
	\right\}
	\right.
\nonumber\\
&&-
	\frac{1}{k'{}^2n(n+4)}
	\left\{
		|k-k'|^{n+2}
		+|k+k'|^{n+2}
\right\}
\nonumber\\
&&+
	\left.
	\frac{k}{k'{}^3n(n+2)(n+4)}
	\left\{
		(k+k')^{n+2}
		-|k-k'|^{n+2}
	\right\}
	\right]\delta(\mbi{k}-\mbi{p})
\nonumber\\
&&\label{eq:T1T1a}
\end{eqnarray}
\subsection{
$\langle T_{[\mathrm{EM:S1}]}T^*_{[\mathrm{EM:S2}]}\rangle
+\langle T_{[\mathrm{EM:S2}]}T^*_{[\mathrm{EM:S1}]}\rangle
$}
Using Eqs.(\ref{tran}), (\ref{ap1}) and (\ref{ap5}), 
the scalar part of $T(\mbi{k})_{[\mathrm{EM:}2]}$ becomes
\begin{eqnarray}
T(\mbi{k},\tau)_{[\mathrm{EM:S}2]}&=&
\hat{k}_i\hat{k}^jT^{i}_{j}(\mbi{k},\tau)_{[\mathrm{EM:}2]}\nonumber\\
 	&=& \frac{1}{4\pi a^{4}}\int\frac{d^3k'}{(2\pi)^3}
\hat{k}_i\hat{k}^jB^{i}(\mbi{k}') B_{j}(\mbi{k}-\mbi{k}') \label{eq:T_EM_2} 
\end{eqnarray}
Thus cross correlations between Eqs.(\ref{eq:T_EM_S1_1}) and (\ref{eq:T_EM_2}) are given by 
\begin{eqnarray}
&&
	\langle
		T_{[\mathrm{EM:S1}]}(\mbi{k})
		T^*_{[\mathrm{EM:S2}]}(\mbi{p})
	\rangle=
\nonumber\\
&&	\frac{1}{2^3(2\pi)^8 a^{8}}\int d^3k'\int d^3p'
	\hat{k}_l\hat{k}^m
 	\langle
 		B^{q}(k') B_{q}(k-k')B^{l*}(p') B^{*}_{m}(p-p')
 	\rangle,
\label{T1T2_1}
 	\nonumber\\
\end{eqnarray}
and
\begin{eqnarray}
&&
	\langle
		T_{[\mathrm{EM:S2}]}(\mbi{k})
		T^*_{[\mathrm{EM:S1}]}(\mbi{p})
	\rangle
=\nonumber\\
&&	\frac{1}{2^3(2\pi)^8 a^{8}}\int d^3k'\int d^3p'
	\hat{k}_i\hat{k}^j
 	\langle
 		B^{i}(k') B_{j}(k-k')B^{q*}(p') B^{*}_{q}(p-p')
	\rangle,
\label{T2T1_1}
\nonumber\\
\end{eqnarray}
respectively.
Using Eq.(\ref{eq:InMPforSangle}),
Eq.(\ref{T1T2_1}) and (\ref{T2T1_1}) become
\begin{eqnarray}
&&	\langle
		T_{[\mathrm{EM:S1}]}(\mbi{k})
		T^*_{[\mathrm{EM:S2}]}(\mbi{p})
	\rangle=
	\frac{1}{2^3(2\pi)^8 a^{8}}
	\left\{
		\frac{(2\pi)^{n+8}}{2k_\lambda^{n+3}}
		\frac{B^2_{\lambda}}{\Gamma\left(\frac{n+3}{2}\right)}
	\right\}^2
	\int d^3k'\int d^3p'
	k'{}^n|\mbi{k}-\mbi{k}'|^n
	\hat{p}_l \hat{p}^m
\nonumber\\	
&&	\times 
	\left\{
		P^{ql}(\mbi{k}')P_{qm}(\mbi{k}-\mbi{k}')\delta(\mbi{k}'-\mbi{p}')
		\delta((\mbi{k}-\mbi{k}')-(\mbi{p}-\mbi{p}'))
		+P^{l}_q(\mbi{k}-\mbi{k}')P^{q}_{m}(\mbi{k}')\delta(\mbi{k}'-(\mbi{p}-\mbi{p}'))
		\delta((\mbi{k}-\mbi{k}')-\mbi{p}')
	\right\}\label{T1T2_2}
\nonumber\\
\end{eqnarray}
and
\begin{eqnarray}
&&
	\langle
		T_{[\mathrm{EM:S2}]}(\mbi{k},\tau)
		T^*_{[\mathrm{EM:S1}]}(\mbi{p},\tau)
	\rangle
=  \frac{1}{2^3(2\pi)^8 a^{8}}
	\left\{
		\frac{(2\pi)^{n+8}}{2k_\lambda^{n+3}}
		\frac{B^2_{\lambda}}{\Gamma\left(\frac{n+3}{2}\right)}
	\right\}^2
	\int d^3k'\int d^3p'
	k'{}^n|\mbi{k}-\mbi{k}'|^n
	\hat{k}_i\hat{k}^j
	P^{in}(\mbi{k}')P_{jq}(\mbi{k}-\mbi{k}')\times\nonumber\\ 
&&	\left\{
		\delta(\mbi{k}'-\mbi{p}')
		\delta((\mbi{k}-\mbi{k}')-(\mbi{p}-\mbi{p}'))
		+\delta(\mbi{k}'-(\mbi{p}-\mbi{p}'))
		\delta((\mbi{k}-\mbi{k}')-\mbi{p}')
	\right\},\label{T2T1_2}
\end{eqnarray}
respectively.
Integrating Eq.(\ref{T1T2_2}) and Eq.(\ref{T2T1_2}) by $p'$,
we obtain following equations,
\begin{eqnarray}
&&
	\langle
		T_{[\mathrm{EM:S1}]}(\mbi{k})
		T^*_{[\mathrm{EM:S2}]}(\mbi{p})
	\rangle
=\nonumber\\
&&	\frac{1}{2^3(2\pi)^8 a^{8}}
	\delta(\mbi{k}-\mbi{p})
	\left\{
		\frac{(2\pi)^{n+8}}{2k_\lambda^{n+3}}
		\frac{B^2_{\lambda}}{\Gamma\left(\frac{n+3}{2}\right)}
	\right\}^2
	\int d^3k'
	k'{}^n|\mbi{k}-\mbi{k}'|^n \hat{p}_l \hat{p}^m 
		\left\{
		P^{ql}(\mbi{k}')P_{qm}(\mbi{k}-\mbi{k}')
		+P^{l}_q(\mbi{k}-\mbi{k}')P^{q}_{m}(\mbi{k}')
	\right\}\label{T1T2_3}\nonumber\\ 
\end{eqnarray}
and
\begin{eqnarray}
&&
	\langle
		T_{[\mathrm{EM:S1}]}(\mbi{k})
		T^*_{[\mathrm{EM:S2}]}(\mbi{k})
	\rangle
=\nonumber\\
&&	\frac{1}{2^3(2\pi)^8 a^{8}}
	2\delta(\mbi{k}-\mbi{p})
	\left\{
		\frac{(2\pi)^{n+8}}{2k_\lambda^{n+3}}
		\frac{B^2_{\lambda}}{\Gamma\left(\frac{n+3}{2}\right)}
	\right\}^2
	\int d^3k'
	k'{}^n|\mbi{k}-\mbi{k}'|^n
	\hat{k}_i \hat{k}^j
	P^{iq}(\mbi{k}')P_{jq}(\mbi{k}-\mbi{k}'),\label{T2T1_3}\nonumber\\
\end{eqnarray}
respectively.
Similary as above the subsection, term containing the unit vector $\hat{k}$ in Eqs.(\ref{T1T2_3}) and (\ref{T2T1_3}) becomes
\begin{eqnarray}
\hat{k}_lP^{ql}(\mbi{k}')\hat{k}_mP^{m}_q(\mbi{k}-\mbi{k}')
	=
		\frac{k'(1-C^2)(k'-kC)}{|\mbi{k}-\mbi{k}'|^2}.\label{TermOfUnitV}
\end{eqnarray}
Substituting Eq.(\ref{TermOfUnitV}) into Eqs.(\ref{T1T2_3}) and (\ref{T2T1_3}),and combining Eqs.(\ref{T1T2_3}) and (\ref{T2T1_3}), we obtain following equation
\begin{eqnarray}
&&	\langle
		T_{[\mathrm{EM:S1}]}(\mbi{k},\tau)
		T^*_{[\mathrm{EM:S2}]}(\mbi{k},\tau)
	\rangle
	+
	\langle
		T_{[\mathrm{EM:S2}]}(\mbi{k},\tau)
		T^*_{[\mathrm{EM:S1}]}(\mbi{k},\tau)
	\rangle
	\nonumber\\
&&	=\frac{1}{2(2\pi)^8 a^{8}}
	\left\{
		\frac{(2\pi)^{n+8}}{2k_\lambda^{n+3}}
		\frac{B^2_{\lambda}}{\Gamma\left(\frac{n+3}{2}\right)}
	\right\}^2
	\int d^3k'k'{}^n|\mbi{k}-\mbi{k}'|^{n-2}
	k'(1-C^2)(k'-kC)\delta(\mbi{k}-\mbi{p}).\label{T1T2_T2T1}
\nonumber\\ 
\end{eqnarray}
Similary, 
integrating Eq.(\ref{eq:T_EM_S1_4}) by $\mathcal{C}$,
after long but straightforward calculation, we can obtain following equation,
\begin{eqnarray}
&&	\langle
		T_{[\mathrm{EM:S1}]}(\mbi{k},\tau)
		T^*_{[\mathrm{EM:S2}]}(\mbi{k},\tau)
	\rangle
	+
	\langle
		T_{[\mathrm{EM:S2}]}(\mbi{k},\tau)
		T^*_{[\mathrm{EM:S1}]}(\mbi{k},\tau)
	\rangle
=\frac{1}{(2\pi)^7 a^{8}}
	\left\{
		\frac{(2\pi)^{n+8}}{2k_\lambda^{n+3}}
		\frac{B^2_{\lambda}}{\Gamma\left(\frac{n+3}{2}\right)}
	\right\}^2
    \delta(\mbi{k}-\mbi{p})\int dk'k'{}^{n+3}
\nonumber\\ 
&&
\left[	
    \frac{1}{(kk')^2n(n+2)}
	\left\{
		(k+k')^{n+3}
	   -(k-k')^{n+3}
	\right\}
\right.
 	-
	\frac{3}{k^2k'{}^3n(n+2)(n+4)}
	\left\{
		(k-k')^{n+4}
		+(k+k')^{n+4}
	\right\}
\nonumber\\ 
&& 	-
	\frac{1}{k^3k'{}^2n(n+2)(n+4)}
	\left\{
		(k+k')^{n+4}
	   -(k-k')^{n+4}
	\right\}
\left.
	+\frac{3}{k^3k'{}^4n(n+2)(n+4)(n+6)}
	\left\{
		(k+k')^{n+6}
	   -(k-k')^{n+6}
	\right\}
\right]
\nonumber\\
\label{eq:T1T2a}
\end{eqnarray}
\subsection{
$\langle T_{[\mathrm{EM:S2}]}T^*_{[\mathrm{EM:S2}]}\rangle
$}
Using Eq.(\ref{eq:T_EM_2} ), 
a two point correlation function of $T_{[\mathrm{EM:S2}]}(\mbi{k})$ becomes
\begin{eqnarray}
\langle 
	T_{[\mathrm{EM:S2}]}(\mbi{k})
	T^*_{[\mathrm{EM:S2}]}(\mbi{p})
\rangle
= 
	\frac{1}{2^2(2\pi)^8 a^{8}}\int d^3k' d^3p'
	\hat{k}_i\hat{k}^j\hat{p}_l\hat{p}^m
	\left\langle
		 B^{i}(k') B_{j}(k-k'){B^{l}}^*(p'){B_{m}}^*(p-p')
	\right\rangle
\label{eq:T_EM_S2_TPC1} 
\end{eqnarray}
Useing Eqs.(\ref{eq:InMPforSangle}) and (\ref{two_point1}),forthremore integling by $p'$, we get
\begin{eqnarray}
&&
\langle 
	T_{[\mathrm{EM:S2}]}(\mbi{k},\tau)
	T^*_{[\mathrm{EM:S2}]}(\mbi{p},\tau)
\rangle\nonumber\\
&& =
	\delta(\mbi{k}-\mbi{p})
	\frac{1}{2^2(2\pi)^8 a^{8}}
	\left\{
		\frac{(2\pi)^{n+8}}{2k_\lambda^{n+3}}
		\frac{B^2_{\lambda}}{\Gamma\left(\frac{n+3}{2}\right)}
	\right\}^2
	\int d^3k'
		k'{}^n|\mbi{k}-\mbi{k}'|^n
	\hat{k}_i\hat{k}^j\hat{p}_l\hat{p}^m
	\{P^{il}(k')P^{jm}(k-k')	+	P^{i}_{m0}(k')P^{l}_{j}(k-k')\}\nonumber\\
\label{eq:T_EM_S2_TPCF3} 
\end{eqnarray}
Similary as the above subsection, 
the term containing the unit vector $\hat{k}$ on Eq.(\ref{eq:T_EM_S2_TPCF3})
 is expressed as
\begin{eqnarray}
\hat{k}_i\hat{k}^j\hat{p}_l\hat{p}^mP^{il}(k')P_{jm}(k-k')
=
\frac{k'{}^2}{|\mbi{k}-\mbi{k}'|^2}
\left(
	1-\mathcal{C}^2
\right)^2.
 \label{UnitOfT2T2V}
\end{eqnarray}
Similary,
substituting Eq.(\ref{UnitOfT2T2V}) for Eq.(\ref{eq:T_EM_S2_TPCF3})
, we obtain following equation
\begin{eqnarray}
&&
\langle 
	T_{[\mathrm{EM:S2}]}(\mbi{k},\tau)
	T^*_{[\mathrm{EM:S2}]}(\mbi{k},\tau)
\rangle\nonumber\\
&& = 
	\frac{1}{2(2\pi)^7 a^{8}}
	\left\{
		\frac{(2\pi)^{n+8}}{2k_\lambda^{n+3}}
		\frac{B^2_{\lambda}}{\Gamma\left(\frac{n+3}{2}\right)}
	\right\}^2
	\int dk'
	\int^1_{-1} d\mathcal{C}
	k'{}^{n+2}|\mbi{k}-\mbi{k}'|^n
	\frac{k'{}^2}{|\mbi{k}-\mbi{k}'|^2}
	\left(
		1-\mathcal{C}^2
	\right)^2.\nonumber\\
\label{eq:T_EM_S2_TPCF4} 
\end{eqnarray}
Similary, 
integrating Eq.(\ref{eq:T_EM_S2_TPCF4}) by $\mathcal{C}$,
after long calculating, we can obtain the following expression,
\begin{eqnarray}
&&
\langle 
	T_{[\mathrm{EM:S2}]}(\mbi{k})
	T^*_{[\mathrm{EM:S2}]}(\mbi{k})
\rangle\nonumber\\
&& = 
	\frac{1}{(2\pi)^7 a^{8}}
	\left\{\frac{(2\pi)^{n+8}}{2k_\lambda^{n+3}}
	\frac{B^2_{\lambda}}{\Gamma\left(\frac{n+3}{2}\right)}\right\}^2
	\int dk'
	k'{}^{n+4}
	\frac{4}{(kk')^3n(n+2)(n+4)}
	\left[\frac{}{}
		\left\{
			(k+k')^{n+4}
			-|k-k'|^{n+4}
		\right\}
	\right.
\nonumber\\
&&-
		\frac{3}{(kk')(n+6)}
		\left\{
			|k-k'|^{n+6}
			+(k+k')^{n+6}
		\right\}
+
	\left.
		\frac{3}{(kk')^2(n+6)(n+8)}
		\left\{
			(k+k')^{n+8}
			-|k-k'|^{n+8}
		\right\}
	\right]
\nonumber\\
\label{eq:T2T2a} 
\end{eqnarray}
\section{Cut Off Scale of PMF}
The photon mean free path $l_{\gamma}$ is  
\begin{eqnarray}
l_{\gamma}&=&\frac{1}{\sigma_T n_e(\tau)}\propto a^3,
\end{eqnarray}
where $n_e$ is the electron number density.
From Refs.\cite{alfven1,subramanian98a}, 
the cut off scale of PMF is defined by 
\begin{eqnarray}
k_C^{-2}(\tau)=
\left\{
		\begin{array}{rl}
			\frac{B^2}{4\pi(\rho+p)}
			\int^{\tau}_{0}d\tau' 
			\frac{l_{\gamma}}{a},
		&	\tau < \tau_\mathrm{dec},\\
			k_C^{-2}(\tau_\mathrm{dec}),
		&	\tau>\tau_\mathrm{dec}
		\end{array}
\right.
\label{eq:CutOff1}
\end{eqnarray}
where
$\tau_\mathrm{dec}$ is the time of decoupling of photons from baryons.
Following Ref.\cite{Mack02}, for a stchastic magnetic field with a power-law power spectrum, the relation between the effective homogeneous field $B$ and $B_\lambda$ is written by 
\begin{eqnarray}
B=B_\lambda\left(\frac{k_C}{k_\lambda}\right)^{\frac{n+3}{2}}.\label{eq:EffctB}
\end{eqnarray}
Subtstituting Eq.(\ref{eq:CutOff1}) into Eq.(\ref{eq:EffctB}),
we obtain 
\begin{eqnarray}
k_C^{-5-n}(\tau)=
\left\{
		\begin{array}{rl}
			\frac{B^2_\lambda k_\lambda^{-n-3}}{4\pi(\rho+p)}
			\int^{\tau}_{0}d\tau' 
			\frac{l_{\gamma}}{a},
			& \tau < \tau_\mathrm{dec} \\
			k_C^{-5-n}(\tau_\mathrm{dec}), & \tau > \tau_\mathrm{dec}
		\end{array}
\right.
	\label{eq:CutOff_F}
\end{eqnarray}
\bibliography{apssamp}

\end{document}